\documentclass[journal=jacsat,manuscript=article,layout=twocolumn]{achemso}

\usepackage{chemformula} 
\usepackage[T1]{fontenc} 
\usepackage{multirow}
\usepackage{hyperref}
\usepackage{graphicx}
\usepackage{academicons}
\usepackage{xcolor}
\newcommand{\orcid}[1]{\href{https://orcid.org/#1}{\hskip2pt\includegraphics[width=9pt]{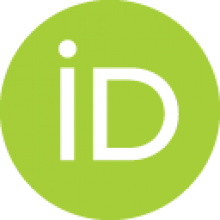}}}

\author {Jessica Perrero \orcid{0000-0003-2161-9120}}
\affiliation{Departament de Qu\'{i}mica, Universitat Aut\`{o}noma de Barcelona, Bellaterra, 08193, Catalonia, Spain}
\alsoaffiliation{Dipartimento di Chimica and Nanostructured Interfaces and Surfaces (NIS) Centre, Universit\`{a} degli Studi di Torino, via P. Giuria 7, 10125, Torino, Italy.}

\author{Joan Enrique-Romero \orcid{0000-0002-2147-7735}}
\affiliation{Univ. Grenoble Alpes, CNRS, Institut de Plan\'{e}tologie et d'Astrophysique de Grenoble (IPAG), 38000 Grenoble, France}
\alsoaffiliation{Departament de Qu\'{i}mica, Universitat Aut\`{o}noma de Barcelona, Bellaterra, 08193, Catalonia, Spain}
\email{juan.enrique-romero@univ-grenoble-alpes.fr}

\author{Berta Mart\'{i}nez-Bachs \orcid{0000-0002-1290-7019}}
\affiliation{Departament de Qu\'{i}mica, Universitat Aut\`{o}noma de Barcelona, Bellaterra, 08193, Catalonia, Spain}

\author{Cecilia Ceccarelli \orcid{0000-0001-9664-6292}}
\affiliation{Univ. Grenoble Alpes, CNRS, Institut de Plan\'{e}tologie et d'Astrophysique de Grenoble (IPAG), 38000 Grenoble, France}

\author{Nadia Balucani \orcid{0000-0001-5121-5683}}
\affiliation{Dipartimento di Chimica, Biologia e Biotecnologie, Universit\`{a} di Perugia, Via Elce di Sotto 8, 06123 Perugia, Italy}
\alsoaffiliation{Univ. Grenoble Alpes, CNRS, Institut de Plan\'{e}tologie et d'Astrophysique de Grenoble (IPAG), 38000 Grenoble, France}
\alsoaffiliation{Osservatorio Astrosico di Arcetri, Largo E. Fermi 5, 50125 Firenze, Italy}

\author{Piero Ugliengo \orcid{0000-0001-8886-9832}}
\affiliation{Dipartimento di Chimica and Nanostructured Interfaces and Surfaces (NIS) Centre, Universit\`{a} degli Studi di Torino, via P. Giuria 7, 10125, Torino, Italy.}

\author{Albert Rimola \orcid{0000-0002-9637-4554}}
\affiliation{Departament de Qu\'{i}mica, Universitat Aut\`{o}noma de Barcelona, Bellaterra, 08193, Catalonia, Spain}
\email{albert.rimola@uab.cat}

\title[An \textsf{achemso} demo]
  {Non-Energetic Formation of Ethanol via CCH Reaction with Interstellar H$_2$O Ices. A Computational Chemistry Study}

\setkeys{acs}{keywords = true}
\abbreviations{iCOMs, ISM, UV, DFT, PES; IRC; ZPE, ASW, BE, BSSE TPD}

\begin{document}

%
%

\onecolumn
\begin{abstract}


Ethanol (CH$_3$CH$_2$OH) is a relatively common molecule, often found in star forming regions.
Recent studies suggest that it could be a parent molecule of several so-called interstellar complex organic molecules (iCOMs).
Yet, the formation route of this species remains debated.
In the present work, we study the formation of ethanol through the reaction of CCH with one H$_2$O molecule belonging to the ice, as a test case to investigate the viability of chemical reactions based on a "radical + ice component" scheme as an alternative mechanism for the synthesis of iCOMs, beyond the usual radical-radical coupling. This has been done by means of DFT calculations adopting two clusters of 18 and 33 water molecules as ice models. Results indicate that CH$_3$CH$_2$OH can potentially be formed by this proposed reaction mechanism. The reaction of CCH with H$_2$O on the water ice clusters can be barrierless (thanks to the help of boundary icy water molecules acting as proton transfer assistants) leading to the formation of vinyl alcohol precursors (H$_2$CCOH and CHCHOH). Subsequent hydrogenation of vinyl alcohol yielding ethanol is the only step presenting a low activation energy barrier. We finally discuss the astrophysical implications of these findings. 
\end{abstract}

\twocolumn

\section{Introduction} \label{sec:intro}

Interstellar complex organic molecules (iCOMs) are compounds between 6-12/13 atoms, in which at least one is carbon, conferring the organic character \cite{Herbst2009,Herbst2017,Ceccarelli_2017}. These molecules are important because i) they can be considered as the simplest organic compounds that are synthesised in space (hence representing the dawn of organic chemistry), and ii) they are the precursors of more complex organic molecules, which can be of biological relevance, such as amino acids, nucleobases and sugars. Indeed, there is robust evidence that the iCOMs formed in the interstellar medium (ISM) were inherited by small objects of the Solar system \cite{Astrochem-Heritage_2012,Ceccarelli_2014,Cazaux_2003,alma-pils2018,Bianchi2019} (e.g., carbonaceous chondrites), which upon alteration mechanisms (e.g., hydrothermal) can be converted into more evolved organic species \cite{Alexander2014,Rotelli2016,Yabuta2007,Callahan2011}, thereby having a potential contribution to the emergence of life on Earth.

The first detection of iCOMs took place in 1971 in massive star formation regions \cite{Rubin1971}, but we had to wait for the beginning of this century for detections in regions that could eventually evolve in Solar-like planetary systems \cite{Ceccarelli2000,Cazaux_2003}. Currently, complex organic molecules have been detected in different astrophysical environments such as prestellar cores, protostellar outflows, protoplanetary disks, and even in external galaxies \cite{Ceccarelli_2017,McGuire2018,Bacman2012,Arce2008,Bianchi2019,Muller2013}.

Two different prevailing paradigms have been postulated for the formation of iCOMs: one based on gas-phase reactions \cite{Charnley1992,Charnley1997,Taquet2016,Balucani2015,skouteris2018,vazart2020}, the other based on radical-radical couplings occurring on grain surfaces \cite{garrod2006, garrod2008}, although other paradigms have been postulated later, like those based on the condensation of atomic C \cite{Ruaud2015, Krasnokutski2020}, excited O-atom insertion \cite{Bergner2017_Oins, Bergner2019}, or the formation of HCO on surfaces as a parent precursor of other iCOMs \cite{Fedoseev2015, Simons2020}. Both prevailing reaction models have the same initial step: formation of hydrogenated species (e.g., CH$_3$OH, NH$_3$, H$_2$CO, CH$_4$) by H-addition on icy grain surfaces in the cold pre-stellar phase. In the gas-phase paradigm, the process follows with the desorption of these species, either thermally when the grain surface reaches ca. 100 K like in hot cores/corinos, non-thermally by photodesorption due to UV incidence on the grains \citep[][]{Bertin2013, Bertin2016, Fayole2011} or by chemical desorption once they are formed \citep[][]{DuleyWilliams1993,Garrod2007,MinissaleDulieu2014,Minissale2016_dust, Takahashi1999b,Vasyunin2013}, or induced by cosmic rays (CR) \citep[][]{Leger1985,1993HasegawaHerbst, Kalvans2016, Dartois2018}. In the gas phase they react with other gaseous species forming iCOMs. In the on-grain paradigm, the hydrogenated species act as parent species of molecular radicals (e.g., CH$_3$, HCO, NH$_2$), formed by the irradiation of UV photons and/or energetic ions, partial hydrogenation, and H-abstraction reactions \citep[][]{garrod2006,garrod2008,Herbst2017, shingledecker_2018, shingledecker_herbst_2018, taquet_multilayer_2012, Minissale_2016_unexpected_CD, jin_garrod_2020} during the cold pre-protostellar stage. Later on, these radicals, due to the warm up of the protostar surroundings (ca. 30K), can diffuse on the icy surfaces and encounter one each other to couple and form the iCOMs. 

These two reaction paradigms have been largely used in combined observational and astrochemical modelling studies to rationalize the presence of iCOMs in the given sources, e.g.,  \cite{Ceccarelli_2017,Codella2017,Ligterink2018,McGuire2017}. And, additionally, they have also been assumed to study the formation of iCOMs by means of quantum chemical simulations \cite{ER2019,ER2021_rrkm,ER2020,Lamberts2019}. By compiling all the available works, it seems that both paradigms are necessary to explain the presence, distribution and abundance of the wide and rich organic chemistry in the ISM. Thus, knowing whether the formation of an iCOM is dominated by surface or gas-phase chemistry is a case-by-case problem. It depends on the nature of the iCOMs and each one has to be addressed as a particular case.

With the different quantum chemical studies addressing on-grain radical-radical couplings, some drawbacks of this paradigm have been identified. One is that the chance for the coupling is a delicate trade-off between the diffusion of the radicals and their desorption. That is, the temperature increase that enables the radical diffusion must be lower than their desorption temperatures. This can give rise to a small temperature window in which the coupling can take place, because this \textit{coupling temperature window} is defined by the lowest diffusion temperature and the lower desorption temperature among the two radicals. For instance, in acetaldehyde formation by the coupling of CH$_3$ and HCO, the coupling temperature window was found to be between the temperatures at which CH$_3$ becomes mobile on the surface (between 9 and 15 K, depending on the adopted diffusion barrier) and the temperature at which the methyl radical would desorb (30 K) \cite{ER2021_rrkm}. Another drawback is that radical-radical couplings are often assumed to be barrierless because they are driven by the coupling of the opposite electronic spins of the radicals. However, the reactions can exhibit energy barriers because the radicals, to proceed with the coupling, need to break the interactions with the icy surfaces. Obviously, as a rule of thumb, the stronger the interaction, the higher the energy barrier. A third limitation of this paradigm is that radical-radical couplings can have competitive channels inhibiting the efficiency of the iCOMs formation. The competitive reactions are based on H-abstractions from one radical to the other. For instance, the CH$_3$ + HCO coupling can give CH$_3$CHO (acetaldehyde as iCOM) but also CH$_4$ + CO (the H-abstraction product).

In a previous work by some of us \cite{Rimola2018}, an alternative on-grain mechanism different from the radical-radical coupling was proposed. It is based on the reaction of a radical (coming from the gas phase or generated by UV irradiation) with neutral, entire components of the ice, i.e., H$_2$O and CO as the most abundant components. In that work, this mechanism was tested to form formamide (NH$_2$CHO) through the reaction of the radical CN with a water molecule belonging to the ice, i.e., CN + H$_2$O(ice) $\longrightarrow$ NH$_2$CO, in which the resulting species can be easily hydrogenated to form formamide. This alternative mechanism overcomes the problem of i) the coupling temperature window (the diffusion of the radical is not needed because it reacts with an abundant ice component, this way increasing the chance of the encountering among the two reactants) and ii) the competitive channel (they \textit{a priori} do not present any other possible reaction). They, however, present energy barriers since the reaction is between a radical and a neutral closed-shell species. 

The present work aims to investigate this alternative on-grain mechanism by simulating with quantum chemical computations the reactivity of the CCH radical with a water molecule of the ice. The goal is to study the possibility to form ethanol (EtOH), with the formation of vinyl alcohol (VA) as intermediate species, through the following reactions:

\begin{equation}
\text{CCH + H$_2$O $\to$ H$_2$CCOH/HCCHOH}
\label{chem_eqn:CCHtoVA*}    
\end{equation}
\begin{equation}
\text{H$_2$CCOH/HCCHOH + H $\to$ H$_2$CCHOH  (VA)}
\label{chem_eqn:VA*toVA}    
\end{equation}
\begin{equation}
\text{H$_2$CCHOH + 2H $\to$ CH$_3$CH$_2$OH  (EtOH)}
\label{chem_eqn:VAtoEtOH}    
\end{equation}

We consider this path towards ethanol because the electronic structure of CCH is isoelectronic with CN, hence exhibiting a similar reactivity with water as shown in Rimola et al. \cite{Rimola2018}.


CCH is one of the first detected interstellar polyatomic molecules \citep{Tucker1974}.
It is a fundamental and common carbon chain (in fact, the simplest) species in the ISM.
It is found in regions near UV sources, the so-called Photon Dominated Regions (PDRs) \citep[e.g.][]{Teyssier_2004_CCH,bouvier_hunting_2020}, in diffuse and translucent clouds \citep[e.g.][]{2000_Lucas_CCH,Gerin2011}, in protostellar objects \cite{Sakai2008,2013_Sakai_Yamamoto}, in the cavities of the protostellar outflows \citep[][]{oya_substellar-mass_2014, zhang_rotation_2018}.
In addition, CCH has been detected in protoplanetary disks \citep[][]{kastner_unbiased_2014, kastner_ring_2015, miotello_bright_2019, Bergner2019, guzman_molecules_2021}
and external galaxies \citep[e.g.][]{meier_spatially_2005}, probably in their PDR regions/skins.
CCH is also abundant towards the lukewarm ($\leq 40$ K)
envelopes of the so-called Warm Carbon Chain Chemistry(WCCC) objects, which are young Class 0/I sources characterized by higher abundances of carbon chains and lower abundances of iCOMs than those observed in hot corinos (see for example ref. \cite{2013_Sakai_Yamamoto}).
CCH appears at scales of few 1000s AU around the protostellar centre at densities of some 10$^6$ cm$^{-3}$ (e.g., ref. \cite{sakai_chemical_2014}). 
Particularly relevant to this work, 
CCH is also relatively abundant in cold molecular clouds \citep[e.g.][]{Friberg1988,Dickens2000} and prestellar cores \cite{2009_Padovani_CCH}.
Typical CCH column densities range from 10$^{12}$--10$^{15}$ cm$^{-2}$ \citep[][]{higuchi_chemical_2018,bouvier_hunting_2020}, with the highest ones in PDRs \cite{bouvier_hunting_2020}. 
In molecular clouds, the CCH column density is around $10^{14}$ cm$^{-2}$, equivalent to abundances of $\sim 10^{-9}-10^{-8}$ \cite{Friberg1988,Dickens2000}.
Similar abundances are found in prestellar cores, as probably CCH resides in the least depleted zone, similar to the molecular clouds \cite{2009_Padovani_CCH,liu2019-cch,Taniguchi2020}.
In summary, CCH is abundant ($\sim 10^{-9}-10^{-8}$) in cold ($\leq 20$ K) regions where the interstellar dust grains are enveloped by icy mantles.
Thus, it is worth to investigate the possibility that it interacts with the water molecules of the ice to form ethanol, following the path described above (reactions 1 to 3).


The reaction of CCH + H$_2$O has been studied as a gas-phase process at high temperatures, both experimentally and theoretically. Seminal experiments by \citet{CCH-water1995} suggested that the outcome of the reaction was not the result of a direct H-abstraction forming HCCH + OH, and hence that authors proposed a mechanism based on, first, an association between the two reactive partners, forming H$_2$C=CHOH or HC=CHOH, followed by an elimination giving rise H$_2$CCO + H and/or HCCH + OH. Subsequent theoretical calculations \cite{CCH-water2001}, however, indicated that among the different reactive channels, the direct H-abstraction was the most kinetically favourable one, in detriment of the association-elimination mechanism. Investigations on this reaction ended with a combined experimental/theoretical study carried out by some of the first paper authors, concluding that the H-abstraction producing HCCH + OH is indeed the most facile chemical reaction \cite{CCH-water2005}.

In contrast, to the best of our knowledge, no experimental works exist on the reactivity of CCH with water ices. However, the addition of OH radicals to C$_2$H$_2$ ices (isoelectronic with CCH + H$_2$O) at temperatures below 20 K, followed by H-additions, has been studied by several authors, resulting with the formation of several products, among them vinyl alcohol and ethanol. This reactivity usually involves an energetic pre-processing of the ice analogues (i.e., irradiation with ions, electrons and photons) to generate the OH radicals. Among these works, we can find: 1) ion radiolysis experiments (MeV protons) that generate mainly CO, CO$_2$, methanol and ethanol \cite{Moore_Hudson_1998}; 2) MeV protons and far UV photon irradiation that yields vinyl alcohol formation \cite{HudsonMoore2003_C2H2}; 3) UV irradiation and proton radiolysis of the ices that form vinyl alcohol, acetaldehyde, ketene and ethanol \cite{Hudson_Loeffler_2013}; 4) ice irradiation with extreme UV photons that leads to the formation of some iCOMs like ethane, methanol and ethanol, together with some simpler species (e.g. CO, CO$_2$ and methane) \cite{wu2002extreme}; and 5) radiolysis of the C$_2$H$_2$:H$_2$O ices with less energetic protons (200 keV) at 17 K that produce several iCOMs like vinyl alcohol, acetaldehyde, ketene and ethanol, and some other species such as C$_2$H$_4$, C$_2$H$_6$, C$_4$H$_2$ and C$_4$H$_4$ \cite{Chuang2021_C2H2_irradH}. In this later work, it was proposed that once vinyl alcohol is formed by the attack of an OH radical to C$_2$H$_2$, two possible situations may take place: either an intramolecular isomerization step forming acetaldehyde or successive H-additions on vinyl alcohol to form ethanol. 
More recently, non-energetic processes have also been explored, in which C$_2$H$_2$:O$_2$ ices exposed to H atoms at 10 K produce most of the products found in Chuang \textit{et al.} \cite{Chuang2021_C2H2_irradH}, including acetaldehyde, vinyl alcohol, ketene and ethanol.
Other experiments indicate that vinyl alcohol and acetaldehyde can also be formed through other chemical reactions. That is, the O addition to C$_2$H$_2$ mainly results in ketene formation \cite{haller_reaction_1962, Hudson_Loeffler_2013, Bergner2019, Zasimov2020}, while the O addition to more saturated hydrocarbons (acetylene, ethane and ethylene) leads to the formation of different iCOMs (ketene, ethanol and acetaldehyde, and acetaldehyde and ethylene oxide, respectively) \cite{Bergner2019, Bennett2005}. 

Ethanol formation has recently received much attention because it has been postulated to be a parent molecule through which other iCOMs can be formed by different gas-phase reactions such as acetaldehyde, glycolaldehyde, formic acid, acetic acid and formaldehyde (the so-called genealogical tree of ethanol, see \cite{skouteris2018,vazart2020,vazart2019}). Because of this significance, this work focuses on the formation of this ancestor molecule following the reactions \ref{chem_eqn:CCHtoVA*}--\ref{chem_eqn:VAtoEtOH} on water ice surfaces by means of quantum chemical simulations to know if they are energetically feasible.

\section{Methodology }\label{sec:methods}

All the calculations were based on the Density Functional Theory (DFT) and run with the {\sc Gaussian16} software  package  \cite{g16}. 
Geometry optimisations and frequency calculations were all performed by combining the DFT methods with the Pople-based 6-311++G(d,p) basis set  \cite{hehre1972,hariharan1973}. These energies were subsequently refined at 6-311++G(2df,2pd) \cite{krishnan1980} level by performing single point energy calculations on the optimised geometries.
In order to identify the DFT method that better describes our systems (and hence to use it for the reaction simulations on the water ice surface models), we carried out a preliminary benchmarking study using the CCH + H$_2$O and CH$_2$CHOH + H gas-phase reactions as models. Five different hybrid DFT methods were used, which were corrected with Grimme's D3 term or, if possible, with the D3(BJ) version \cite{D2-grimme2006,D3-grimme2010,d3bj_grimme} to account for dispersion interactions. The tested DFT-D methods were: BHLYP-D3(BJ) \cite{bhandhlyp-becke1993,lee1988}, M062X-D3 \cite{m062x-zhao}, MPWB1K-D3(BJ) \cite{mpwb1k-zhao}, PW6B95-D3(BJ) \cite{pw6b95-zhao} and $\omega$B97X-D3 \cite{wb97x-chai}. By comparing the results with those obtained with single energy points at the CCSD(T)/aug-cc-PVTZ \cite{ccsd(t)} level of theory, known as the "gold-standard" in quantum chemistry, we found that the $\omega$B97X-D3 method showed the best performance when modeling the water addition to CCH, while MPWB1K-D3(BJ) described better the hydrogenation of CH$_2$CHOH  (see \S~\nameref{sec:results} below). Accordingly, the CCH + H$_2$O and the CH$_2$CHOH + H reactions on the water ice cluster models were computed, respectively, at the $\omega$B97X-D3/6-311++G(2df,2pd)//$\omega$B97X-D3/6-311++G(d,p) and the MPWB1K-D3(BJ)/6-311++G(2df,2pd)//MPWB1K-D3(BJ)/6-311++G(d,p) theory levels.

All the stationary points of the potential energy surfaces (PESs) were characterized by their analytical calculation of the harmonic frequencies as minima (reactants, products, and intermediates) and saddle points (transition states).  When needed, intrinsic reaction coordinate (IRC) calculations at the level of theory adopted in the geometry optimizations were carried  out to  ensure  that  a given  transition state actually connects  with the  corresponding  minima. Thermochemical corrections to the potential energy values were carried  out  using  the  standard  rigid  rotor/harmonic oscillator formulas to compute the zero point energy (ZPE) corrections \cite{mcquarrie}. Since the systems are open-shell in nature, calculations were performed within the unrestricted formalism.

We additionally calculated the tunnelling crossover temperatures, i.e., the temperature below which quantum tunnelling becomes the main mechanism for trespassing the potential energy barriers. To this end, we used Eq. \ref{eq:crossover_T} \citep[][]{FermannAuerbach2000_Tc}, where $\Delta H^{\ddagger}$ is the ZPE-corrected barrier height, $\omega^{\ddagger}$ is the frequency associated to the TS, and $k_B$ and  $\hbar$ are the Boltzmann's and reduced Planck's constants. This temperature indicates what reactions may actually have an important role at low temperatures despite of having an activation barrier.

\begin{equation}
    T_c = \frac{\hbar\omega^{\ddagger}\Delta H^{\ddagger}/k_B}{2\pi \Delta H^{\ddagger}-\hbar\omega^{\ddagger}\ln(2)}
    \label{eq:crossover_T}
\end{equation}

\begin{figure*}[!htb]
    \centering
    \includegraphics[width=0.8\textwidth]{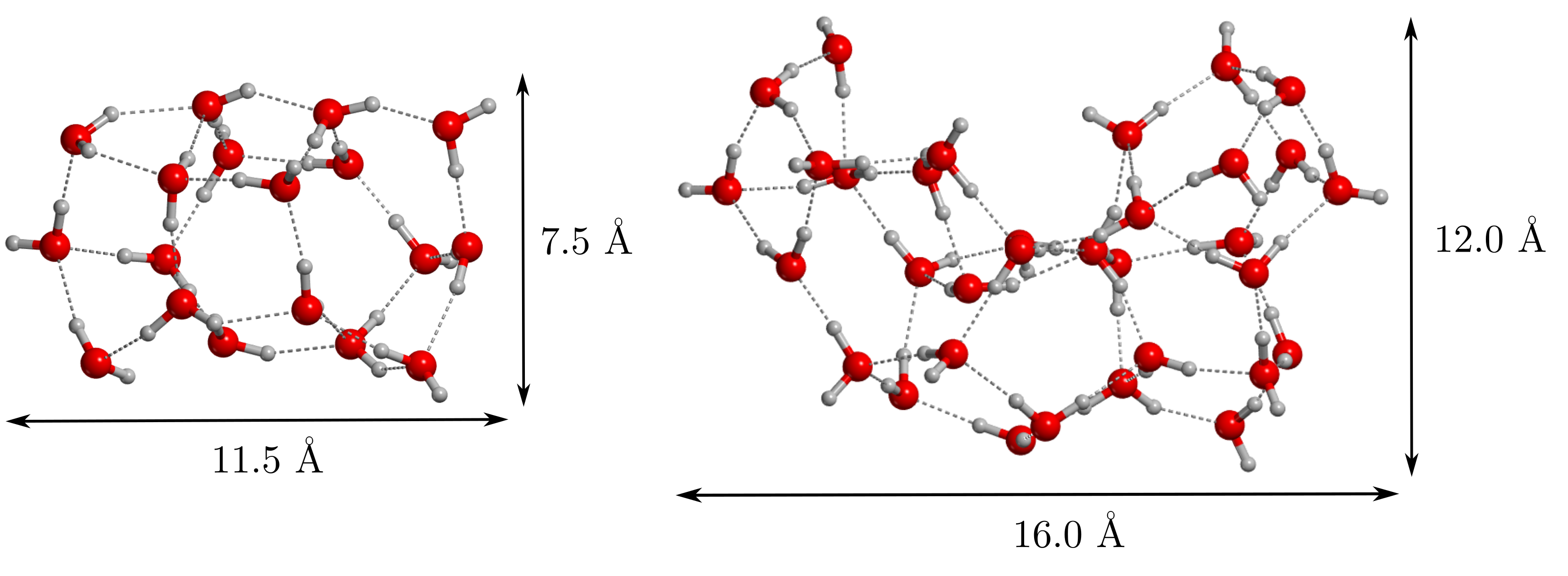}
    \caption{Structures of the 18 and 33 water molecules clusters, optimized at the $\omega$B97X-D3/6-311++G(d,p) level of theory.}
    \label{fig:water_models}
\end{figure*}

The surfaces of amorphous solid water (ASW) ice coating interstellar grains were simulated by two different cluster models: one consisting of 18 water molecules, the other of 33 water molecules (hereafter referred to as W18 and W33, see Figure \ref{fig:water_models}). While the former represents a compact, flat water ice surface, the latter presents a cavity structure of about 6 \r{A}. For more details, we refer the reader to see \cite{rimola2014,Rimola2018,ER2019}.\\

For the calculation of the binding energies (BEs) of CCH interacting with the H$_2$O ice surface models (i.e., CCH/surf complexes) we adopted the same electronic structure methodology as for reactivity, namely, for each CCH/surf complex and corresponding isolated components, geometry optimizations and frequency calculations (and hence ZPE corrections) were computed at $\omega$B97x-D3/6-311++G(d,p), followed by single point energy calculations at the improved $\omega$B97x-D3/6-311++G(2df,2pd) level. 
Basis set superposition error (BSSE) was corrected following the Boys and Bernardi counterpoise method. The final, corrected, adsorption energy ($\Delta E_{ads}^{CP}$) was calculated as:
\begin{align*}
    \Delta E_{ads}^{CP} (CCH/surf) =  \Delta E_{ads} + {BSSE}(CCH) \\
    + {BSSE}(surf) + \Delta ZPE
     \label{eq:binding_bsse}
\end{align*}
where $\Delta E_{ads} = E(CCH/surf) - E(CCH) - E(surf)$ refers to the BSSE-non-corrected adsorption energy. 
Note that we used the same sign convention as in \cite{ER2019}, namely, the adsorption energy is the negative of the binding energy: $\Delta E_{ads}^{CP} = - BE$.

\section{Results}\label{sec:results}

\subsection{Benchmarking study}

\begin{figure*}[!htb]
    \centering
    \includegraphics[width=0.8\textwidth]{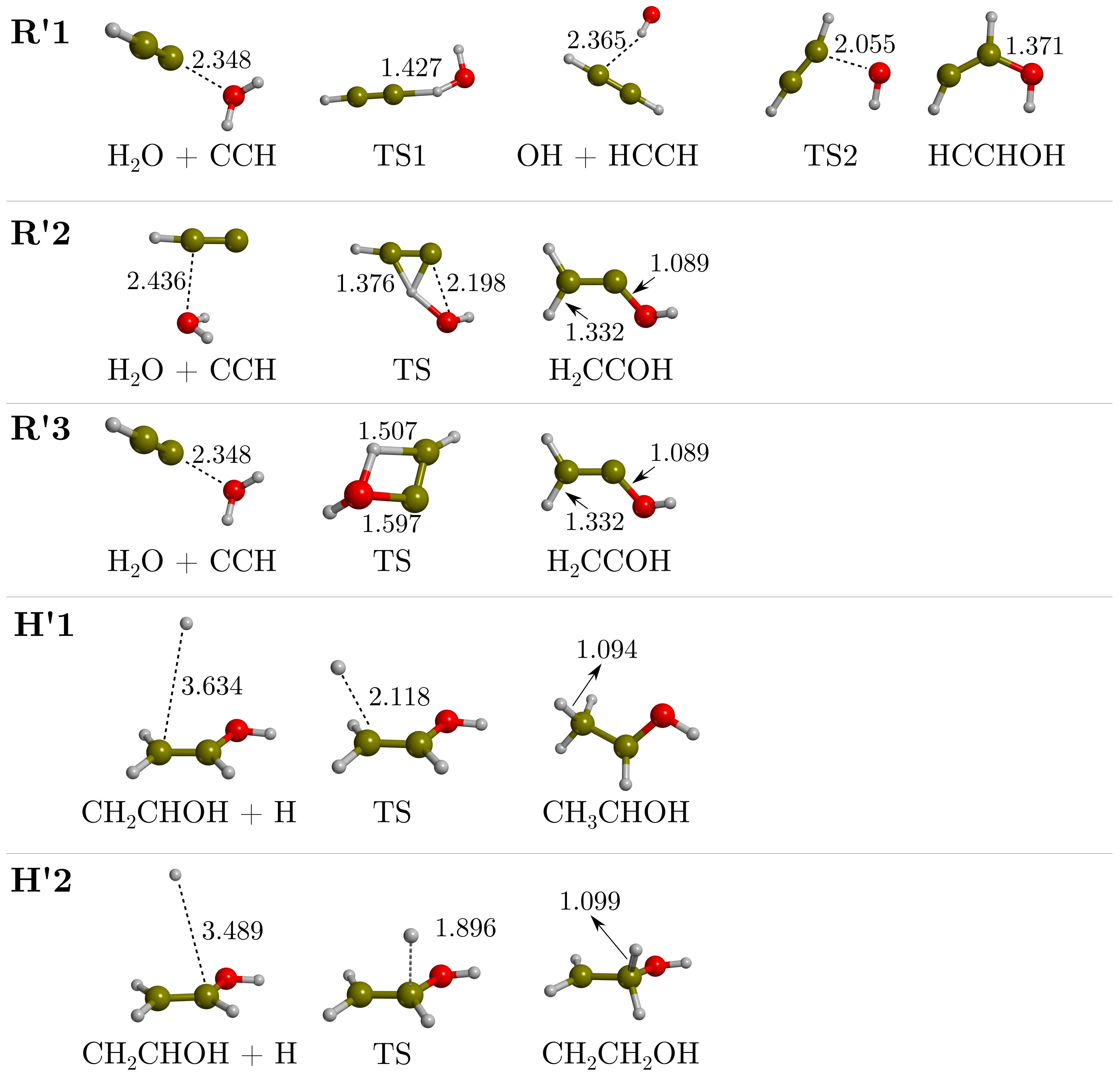} \\
    \caption{Stationary points identified in the benchmarking study. Reaction pathways \textbf{R'1}, \textbf{R'2} and \textbf{R'3} refer to gas-phase CCH + H$_2$O reaction and are optimized at $\omega$B97X-D3. \textbf{H'1} and \textbf{H'2} correspond to the two hydrogenation pathways of vinyl alcohol, optimized at MPWB1K-D3(BJ). Distances are in \r{A}.}
    \label{fig:benchmark_pes}
\end{figure*}

As mentioned above, we carried out a preliminary benchmarking analysis for the reactivity using two gas-phase model reactions, CCH + H$_2$O and CH$_2$CHOH + H, to find the DFT method that describes better the reaction properties. For CCH + H$_2$O, we found three possible reaction pathways, labelled as \textbf{R'1}, \textbf{R'2} and \textbf{R'3}, the stationary points of which being shown in Figure \ref{fig:benchmark_pes}.


\textbf{R'1} follows a stepwise mechanism, the first step involving the formation of acetylene (HCCH) and the hydroxyl radical (OH) as intermediate species, and the second step consisting of the condensation of these intermediates to form the HCCHOH radical. In contrast, both \textbf{R'2} and \textbf{R'3} adopt a concerted mechanism. These two reactions involve a C-O bond formation followed by a H-transfer to the other C atom to form the H$_2$CCOH radical (isomer of HCCHOH, the product of \textbf{R'1}). The difference between \textbf{R'2} and \textbf{R'3} is that, in the former, the H-transfer comes first, followed then by a spontaneous C-O bond formation, while in the latter the C-O bond formation and the proton transfer evolve in a synchronized way.


The computed energetics of these reaction pathways using the different quantum chemical methods are shown in Table \ref{tab:benchmark}. The optimized geometries of the stationary points are available in the supporting information (SI). As it can be seen, the overall best functional for these reactions of water addition to CCH is $\omega$B97X-D3, with an average unsigned error of $\sim$5\% in the energy barriers and $\sim$6\% in the whole energetics. The performance of the other functionals is, from better to worse: MPWB1K-D3(BJ), PW6B95-D3(BJ), BHLYP-D3(BJ) and M062X-D3. 

\begin{table*}[!htb]
\centering
\caption{Relative energies (in kJ mol$^{-1}$) of the stationary points involved in the reaction pathways found for the gas-phase CCH + H$_2$O reaction model (\textbf{R'1}, \textbf{R'2} and \textbf{R'3}) and for the gas phase hydrogenation (\textbf{H'1} and \textbf{H'2}) for all the DFT-D methods and the CCSD(T) method used for the benchmarking study. See the Methodology section for more details. Rows indicated as ``TS \% RX'' show the unsigned error (in percentage) relative to the predicted energy of the TS.
Rows indicated as ``Avg \% RX'' show the averaged unsigned errors (in percentage and accounting for all the stationary points) of each DFT-D method with with respect to the CCSD(T)//$\omega$B97X-D3 for R'1, R'2 and R'3 and CCDS(T)//MPWB1K-D3(BJ) for H'1 and H'2 reference values. The last row indicates the global averaged unsigned error (in percentage).}

\label{tab:benchmark}
\resizebox{\textwidth}{!}{%
\begin{tabular}{|c|l|cccccc|}
\hline
Reaction            & Step           & BHLYP-D3(BJ)  & M062X-D3  & MPWB1K-D3(BJ) & PW6B95-D3(BJ) & $\omega$B97X-D3 & CCSD(T) \\
\hline \hline
\multirow{5}{*}{R'1} & H$_2$O + CCH   & 0.0    & 0.0    & 0.0    & 0.0    & 0.0          & 0.0     \\
                    & TS1            & 28.5   & 8.9    & 25.6   & 19.6   & 29.0         & 29.1    \\
                    & OH + HCCH      & -91.6  & -57.0  & -72.5  & -63.0  & -62.3        & -69.0   \\
                    & TS2            & -68.5  & -62.8  & -60.2  & -61.1  & -53.5        & -51.5   \\
                    & HCCHOH         & -217.0 & -222.0 & -217.7 & -203.6 & -205.9       & -193.8  \\
\hline
\multicolumn{2}{|c|}{TS1 \%R'1}        & 2.1  & 69.5  & 11.8 & 32.6 & 0.4        &     - -    \\
\multicolumn{2}{|c|}{TS2 \% R'1}        & 33.2  & 22.1  & 17.0 & 18.8 & 4.0        &     - -    \\
\multicolumn{2}{|c|}{Avg \% R'1}        & 20.0 & 30.9 & 11.5 & 16.3 & 5.1         &     - -    \\
\hline \hline
\multirow{3}{*}{R'2} & H$_2$O + CCH      & 0.0    & 0.0    & 0.0 & 0.0 & 0.0       & 0.0   \\
                     & TS1   & 117.4  & 107.5  & 96.4 & 90.8 & 108.0       & 99.5   \\
                     & H$_2$CCOH         & -249.3 & -246.8 & -266.3 & -253.2 & -247.8       & -241.11  \\
\hline
\multicolumn{2}{|c|}{TS \% R'2}        & 18.0  & 8.1  & 3.1 & 8.7 &  8.5       &     - -    \\
\multicolumn{2}{|c|}{Avg \% R'2}          & 10.7     & 5.2   & 6.8   & 6.9   & 5.6 &    - -     \\
\hline \hline
\multirow{3}{*}{R'3} & H$_2$O + CCH & 0.0    & 0.0    & 0.0    & 0.0    & 0.0          & 0.0     \\
                    & TS             & 120.3   & 87.7   & 98.6   & 123.5   & 113.0         & 105.4    \\
                    & H$_2$CCOH      & -240.3 & -244.8 & -241.9 & -228.8 & -229.7       & -214.1  \\
\hline 
\multicolumn{2}{|c|}{TS \% R'3}        &  16.7 & 16.8 & 6.5 & 17.1 & 7.2        &     - -    \\
\multicolumn{2}{|c|}{Avg \% R'3}        &   14.5 & 15.6 & 9.7 & 12.0 & 7.2        &     - -    \\
\hline\hline
\multicolumn{2}{|c|}{Global Avg \%}    & 15.1 & 17.2 & 9.3 & 11.7 & 6.0                   &  - -\\ 
\hline
\hline
\multirow{3}{*}{H'1} & CH$_2$CHOH + H & 0.0    & 0.0    & 0.0    & 0.0    & 0.0          & 0.0     \\
                    & TS             & 0.3   & -   & 4.1   &  2.3  &   11.4       & 6.5    \\
                    & CH$_3$CHOH      &  -192.6 & - & -183.3 & -177.1 &  -188.1      & -176.0  \\
\hline
\multicolumn{2}{|c|}{TS \% H'1}        & 116.4  & -  & 43.6 & 78.5 &    93.0     &     - -    \\
\multicolumn{2}{|c|}{Avg \% H'1}          &  62.9   & -  & 23.9   &   39.6 & 49.9 &    - -     \\
\hline
\multirow{3}{*}{H'2} & CH$_2$CHOH + H & 0.0    & 0.0    & 0.0    & 0.0    & 0.0          & 0.0     \\
                    & TS             & 8.0   & -   & 14.2   & 12.5   & 21.6         & 15.9    \\
                    & CH$_2$CH$_2$OH      & -157.9 & - & -143.0 & -177.1 & -149.0       & -142.0  \\
\hline
\multicolumn{2}{|c|}{TS \% H'2}        & 54.2  & - & 11.9  &  23.7 &    38.5     &     - -    \\
\multicolumn{2}{|c|}{Avg \% H'2}          &  32.7    & - & 6.3   & 24.1   & 21.7 &    - -     \\
\hline
\hline
\multicolumn{2}{|c|}{Global Avg \%}    & 47.8   & -   & 15.1  & 31.8   &     35.8                &  - -\\ 
\hline
\end{tabular} %
}
\end{table*}

For the hydrogenation steps (reactions \ref{chem_eqn:VA*toVA}--\ref{chem_eqn:VAtoEtOH}), we only considered the hydrogenation of vinyl alcohol, namely, CH$_2$CHOH + H, because it is the unique step that can exhibit an energy barrier due to involving a closed-shell molecule (vinyl alcohol) with  a radical (H atom). In contrast, the other steps consist in the hydrogenation of radical species, which are barrierless processes due to the spontaneous spin-spin coupling. For this H-addition reaction, we found two possible pathways (labelled as \textbf{H'1} and \textbf{H'2} in Figure \ref{fig:benchmark_pes}), leading to two different products (CH$_3$CHOH and CH$_2$CH$_2$OH, respectively), depending on which C atom the H addition takes place. 
\textbf{H'1} and \textbf{H'2} share a similar mechanism, in which the H atom in the reactant structures is located at ca. 3.5 \r{A} from the C atom with which it will react. Results show that path \textbf{H'1} is more favorable than \textbf{H'2}, both for the stability of the product and for presenting a lower activation energy barrier (see Table \ref{tab:benchmark}). This is because the \textbf{H'1} product (CH$_3$CHOH) exhibits a better delocalization of the unpaired electron with respect to the \textbf{H'2} product (CH$_2$CH$_2$OH). Among the used DFT methods, the functional with the smallest average unsigned error compared to CCSD(T) single point energy calculations is MPWB1K-D3(BJ). The performance of the other functionals, from better to worse, is: PW6B95-D3(BJ), $\omega$B97X-D3 and BHLYP-D3(BJ). M062X-D3 was discarded for convergence problems of the reactant structures. 
Therefore, according to this benchmarking study, the $\omega$B97X-D3 DFT method has been chosen to simulate the addition of water to CCH on the W18 and W33 cluster models and MPWB1K-D3(BJ) has been adopted for the hydrogenation step of vinyl alcohol.

\subsection{Adsorption of CCH on water ice and binding energies}

The complexes formed when CCH adsorbs on W18 and W33 are shown in the R structures of Figures \ref{fig:RX_CCHonW18} and \ref{fig:RX_CCHonW33}, respectively. Table \ref{tab:be_w18} reports the computed binding energies and the different contributions, as detailed in \S~\nameref{sec:methods}. 

In most of the cases, a nonclassical hemibond between the CCH species and a H$_2$O of the ice is established, due to the formation of a two-center three-electron system between the unpaired electron of CCH and a lone pair of H$_2$O. This interaction is highlighted by the computed spin density values and maps, which clearly indicate a delocalization of the unpaired electron between the two centers (see data in SI). The distances of these hemibonds vary between 2.1--2.3 \r{A} (see the R structures of the \textbf{R2}, \textbf{R3}, \textbf{R2-1} and \textbf{R2-2} sequences of Figures \ref{fig:RX_CCHonW18} and \ref{fig:RX_CCHonW33}). The only exception not presenting an hemibonded complex is the structure R of \textbf{R1} (Figure \ref{fig:RX_CCHonW18}), in which classical, weak hydrogen bond (H-bond) interactions are established between CCH and the W18 ice model. Interestingly, any attempt to find a pure H-bonded complex (i.e., without any hemibonded interaction) on W33 failed, the geometry optimizations collapsing into the hemibonded structures. This reinforces the idea that the more surface water molecules, the higher the possibility to form hemibonded complexes, although most of the hemibonded complexes also present H-bond interactions. This is because the outermost water molecules of the cluster are unsaturated in terms of H-bonds, presenting H- and O-dangling atoms ready to establish H-bond and hemibonded interactions, respectively.
 Remarkably, hemibonded systems also form in the reactant structures of the gas-phase reactions (see above). However, these gas-phase systems present a lower hemibonded character than the complexes on W18 and W33 because the spin is less delocalized between the two centers (see spin density values and maps in SI).

As expected, hemibonded complexes present larger BE values than H-bonded ones (see Table \ref{tab:be_w18}). The difference in the BEs between the complexes shown in \textbf{R2} and \textbf{R3} (i.e., on W18) arises from the orbital occupied by the CCH unpaired electron. In the former, the unpaired electron belongs to the $\pi$ system of the CCH (the spin density is shared between the two C atoms and the linearity of CCH becomes broken upon hemibond formation), while in the latter to a $\sigma$ orbital of the C-end of CCH, in both cases interacting with a lone pair of H$_2$O to form the hemibond. As the orbital overlap is more efficient in $\sigma$ orbitals than in $\pi$ ones, the latter complex presents a larger BE than the former (49.9 and 37.9 kJ mol$^{-1}$, respectively). On W33, both hemibonded complexes have similar BE values (89.7 and 86.0 kJ mol$^{-1}$) as in both systems the unpaired electron occupies a $\pi$ orbital. Remarkably, the values on W33 are notably higher than those on W18, and we investigated which might be the cause. We checked for a correlation between the BE and the number of water molecules forming the cluster, carving several structures from our W33 model, that is, by removing water molecules from the edges of the model in order to built a set of water clusters with decreasing dimensions (and blocking some O atoms of the cluster to prevent the cavity from collapsing). By proceeding this way, we found evidence that the increasing number of interactions between CCH and the ice, together with the cooperativity of H-bonds, is responsible for the increment of the BE. Data of this analysis \textbf{are} provided in SI.

Finally, it is worth mentioning that we performed a preliminary benchmarking study on the binding energies of the dimeric CCH/H$_2$O system. Results (available in SI) indicate that while the H-bonded dimer is very well described by most of the DFT methods, this is not the case for the hemibonded one. However, for the particular case of $\omega$B97X-D3, the computed binding energies are somewhat overestimated, belonging to the group of the functionals that better compare with CCSD(T). Accordingly, the computed binding energies of the hemibonded complexes arising from CCH in interaction with W18 and W33 should be considered overestimated by some amount. Despite this drawback, we would like to stress that the main scope of the work is on the reactivity between CCH and H$_2$O followed by H-additions and that the used DFT methods are actually accurate enough for this purpose, as shown above. In this case, the error in the binding energies self-cancelled when deriving the energy features (energy barriers and reaction energies) of the reactions.


\begin{table*}[!htb]
\centering
\caption{Binding energy (BE) values (in kJ mol$^{-1}$) of CCH on the W18 and W33 cluster models according to the computed complexes shown in Figures \ref{fig:RX_CCHonW18} and \ref{fig:RX_CCHonW33} (the R structures of the reactions (Rx) \textbf{R1}, \textbf{R2}, \textbf{R3}, \textbf{R2-1} and \textbf{R2-2}). The contributions from the pure potential energy values ($\Delta E_{ads}$), the dispersion corrections ($\Delta E_{disp}$), the zero point energy corrections ($\Delta ZPE$), and the BSSE corrections ($\Delta BSSE$) are also shown.}
\label{tab:be_w18}
\begin{tabular}{|c|l|cccc|c|}
\hline
Surface & {Rx}          & $\Delta E_{ads}$ & $\Delta E_{disp}$ & $\Delta ZPE$ & $\Delta BSSE$ & BE \\
\hline
\multirow{3}{*}{W18}& R1 &  -21.2  &  -3.3  &  -0.6 & 1.4 & 23.7 \\
& R2 & -32.8    & -10.1  & 3.0     & 2.1  & 37.9   \\
& R3         & -32.2  & -22.1  & 2.2  & 2.2  & 49.9    \\
\hline
\multirow{2}{*}{W33}& R2-1 & -97.1 & -7.2 & 11.2 & 3.3 & 89.7\\
 & R2-2 & -81.6 & -19.0 & 11.2 & 3.5 & 86.0 \\
\hline
\end{tabular}%
\end{table*}

\subsection{Reactivity between CCH and H$_2$O on the ice surface models}

\subsubsection{On the W18 ASW ice cluster model}\

Following the three reaction types found in the benchmark study, we  tried to reproduce them on top of the W18 ice cluster model. However, significant differences in the mechanisms have been found, precisely because they occur on the ice model. Indeed, the larger number of water molecules infers that i) there are no concerted mechanisms to work, and ii) water-assisted proton-transfer reactions are operative. 
Water-assisted proton transfer mechanisms have been elucidated theoretically in the latest of the last century\cite{Chalk-1997} and can induce important decrease in the energy barriers with respect to the non-water-assisted analogues. This is because the assisting water molecules bridge the accepting/donating proton processes with their neighboring molecules and, at the same time, reduce the strain of the rings in the transition state structures. Hence that this mechanism is also called proton-transport catalysis. Such catalytic effects have also been observed in processes of interstellar interest on icy surfaces \cite{Rimola2018,Molpeceres-2021,Rimola-glycine}. It is worth mentioning that the water-assisted proton-transfers, to be catalytically effective, need a proper H-bond connectivity among the bridging water molecules, from the first to the last proton transfer. This means that, if the H-bonding network is truncated due to the presence of interstitial unpurities, the mechanism is not operative. This is an aspect not to overlook in interstellar ices as they contain minor volatile species in the ices that can obstruct the chained proton relays. Remarkably, it is worth stressing that the catalytic transfers involve H atoms with a proton character and not atomic radicals. This is because the H species exchanged during the transfers are proton-like atoms chemically bonded to a more electronegative O atom. 

Since the identified reaction channels on W18 differ significantly from the gas-phase ones, we redefine the  \textbf{R'1--3} model channels into \textbf{R1--3}, which adopt the following simplified mechanistic steps:


\begin{description}
\item[R1] H$_2$O + CCH $\to$ OH + HCCH $\to$ HCCHOH, 
\item[R2] H$_2$O + CCH $\to$ CC(H)--OH$_2$ $\to$ HCCHOH,
\item[R3] H$_2$O \textbf{+ CCH} $\to$ H$_2$O--CCH $\to$ H$_2$CCOH.
\end{description}

\begin{figure*}[!htbp]
    \centering
    \begin{tabular}{@{} c}
    \includegraphics[width=0.6\textwidth]{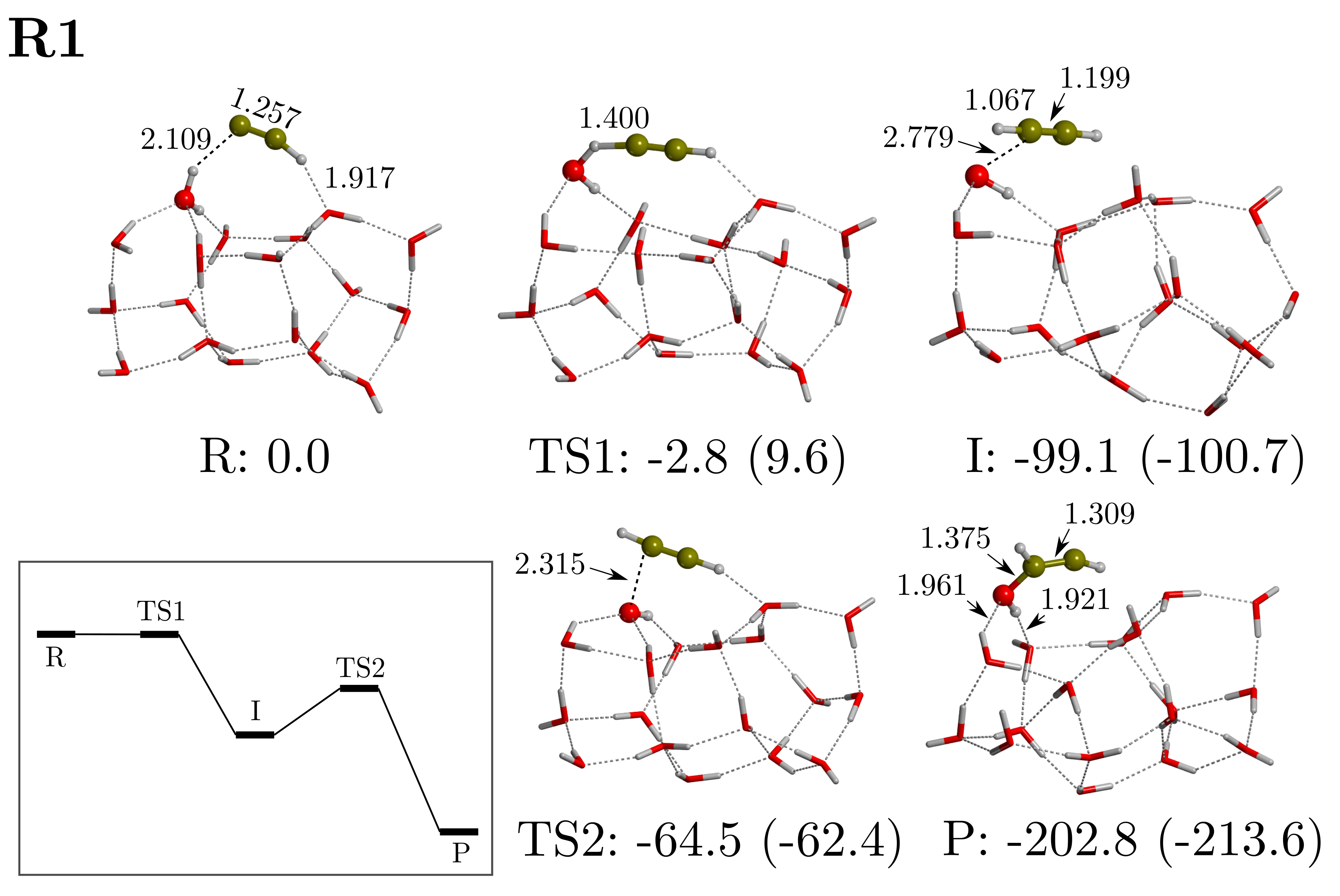} \\
    \hline
    \includegraphics[width=0.6\textwidth]{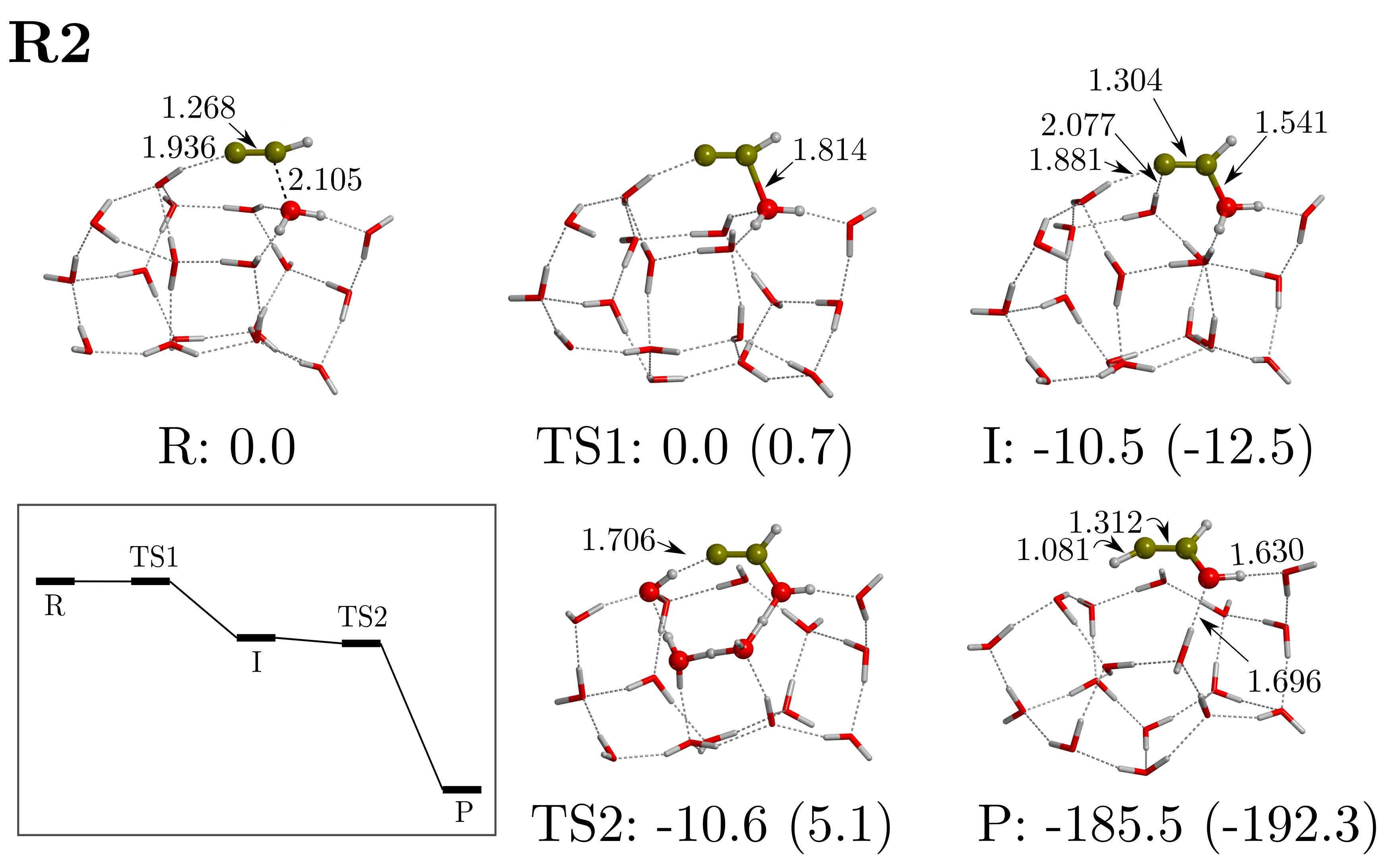}\\
    \hline
    \includegraphics[width=0.6\textwidth]{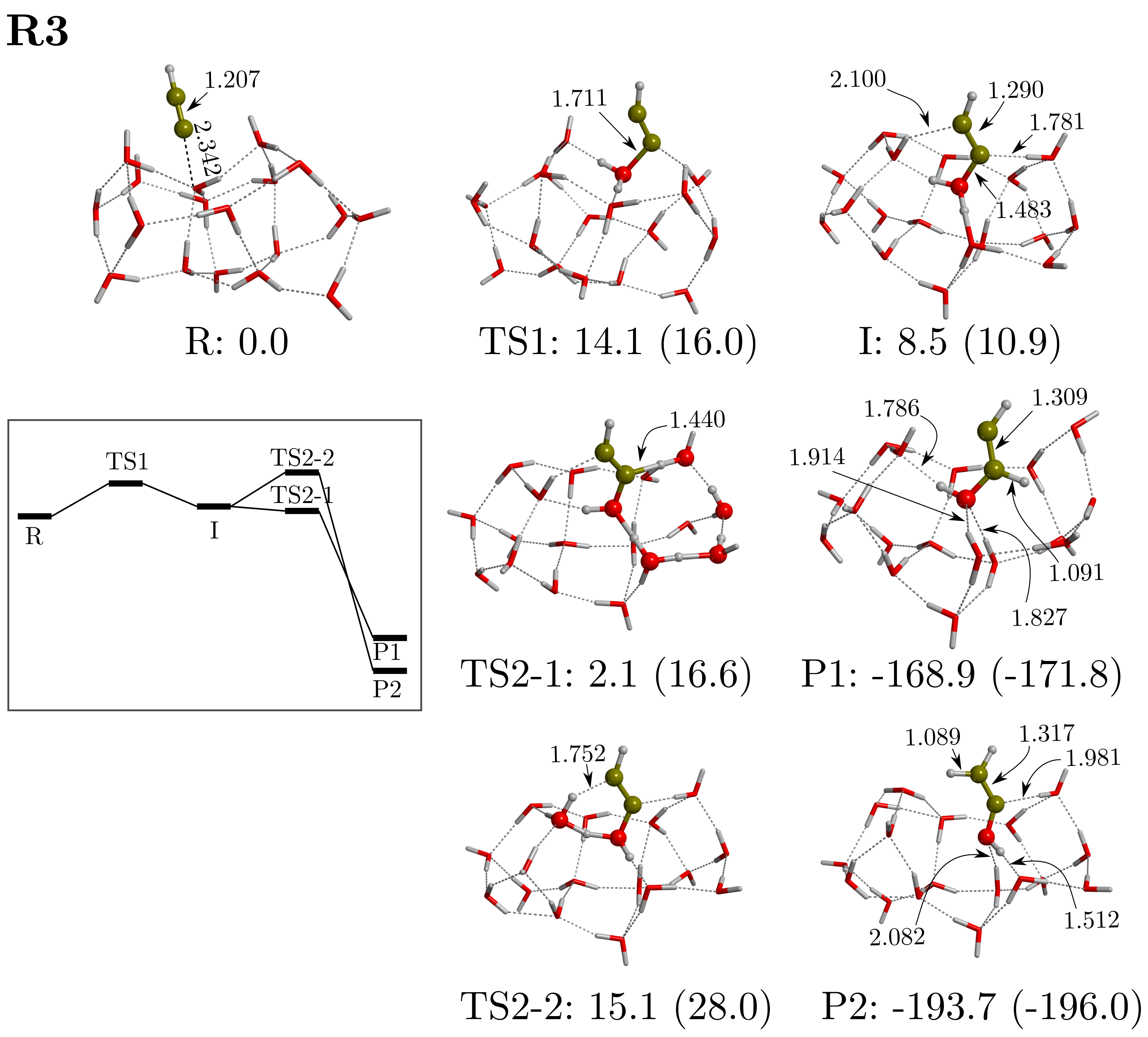}\\
    \end{tabular}
    \caption{\scriptsize Computed potential energy surfaces (PESs) of the reactions \textbf{R1--3} between CCH and the W18 ASW ice cluster model. Bare energy values correspond to relative ZPE-corrected values, while values in parenthesis to those missing this correction. The miniature panels sketch the ZPE-corrected PESs. Energy units are in kJ mol$^{-1}$ and distances in \r{A}.}
    
    \label{fig:RX_CCHonW18}
\end{figure*}

The stationary points and the energy profiles of these reaction pathways are shown in Figure \ref{fig:RX_CCHonW18}. 

The \textbf{R1} path remains the same as the \textbf{R'1} one, namely, formation of HCCH and OH as intermediate species by H-transfer from the reactive H$_2$O molecule to CCH, and then coupling of the OH to HCCH to form the HCCHOH radical as vinyl alcohol precursor. 
Both \textbf{R2} and \textbf{R3} starts with the formation of the hemibonded systems (see SI for their spin densities). 
However, since the involved C atoms are different, the pathways proceed in a different way as well. The hemibonded structure of \textbf{R2} evolves towards the formation of the CC(H)-OH$_2$ intermediate, which is followed by a proton transfer, from the OH$_2$ moiety and adopting a water-assisted proton-transfer mechanism, to form the final HCCHOH radical. In contrast, the hemibonded structure of \textbf{R3} transforms into H$_2$O-CCH as the intermediate species. From this intermediate, two different paths are possible, namely, from the OH$_2$ moiety, a proton transfer to i) the central C atom (TS2-1) or ii) the terminal C atom (TS2-2), forming HCCHOH or H$_2$CCOH, respectively. Interestingly, both paths proceed through a water-assisted proton-transfer mechanism. The fact that \textbf{R3} exhibit these two paths is because there are well oriented H-bonds in the water cluster allowing for the water-assisted mechanism in the two directions, while this is not possible in \textbf{R2} due to the geometry of the intermediate structure.

In relation to the energetics (see also Figure \ref{fig:RX_CCHonW18}), the first steps (TS1) of paths \textbf{R1} and \textbf{R2} become submerged  below the energy of reactants once they are ZPE-corrected. The same happens for the second step (TS2) of path \textbf{R2}. This means that \textbf{R2} is an overall effectively barrierless reaction on W18. In contrast, \textbf{R1} has a low energy intermediate (associated with the formation of HCCH) and, consequently, the second step (TS2) encounters a relatively high energy barrier of 34.6 kJ mol$^{-1}$ (with respect to the intermediate). For \textbf{R3}, the first step already presents a non-negligible energy barrier, even when it is ZPE-corrected (14.1 kJ mol$^{-1}$). For the second steps, TS2-1 has a lower energy than that of TS2-2 because the geometry associated with the latter saddle point is more strained, as the cycle created by the water assistant molecules is smaller. All the reaction paths present negative and very large reaction energies, indicating that the formed vinyl alcohol radicals are more stable than the reactant states. 
We observed that, in some cases, the ZPE corrections added to the potential energies are very large (see, for instance, TS1 in \textbf{R1}, TS2 in \textbf{R2}, and TS2-1 and TS2-2 in \textbf{R3}). Interestingly, these transition states involve the motion of H atoms of the water molecules, either as a direct proton-transfer (case of TS1 in \textbf{R1}) or as a water-assisted proton-transfer mechanism (cases of TS2 in \textbf{R2}, and TS2-1 and TS2-2 in \textbf{R3}). In contrast, when a C-O bond is formed (namely, without involving any proton motion) it results in a slight ZPE-correction. In order to understand this behaviour, we simulated the IR spectra of R, TS1, I and TS2-1 of \textbf{R3} (see SI). The spectra of R and TS1, where only the C--O bond forms, show mostly common features, with the exception of a couple of bands around 3000 cm$^{-1}$, due to O-H stretching of the water molecules surrounding the CCH. On the other hand, comparing the spectrum of TS2-1 with R highlights the presence of a number of features at a shorter wavenumber, arising from the water-assisted proton-transfer. Given that the ZPE arises from the sum of the energy of all the vibrational modes, the shorter wavelength features of TS2-1 explain why its ZPE is smaller than R and hence that the resulting $\Delta$ZPE is negative.

As a final comment, we would like to stress the influence of the saturated state (in terms of H-bonding) of the reacting water molecule in the energy barriers. Indeed, according to the potential energy values, the TS1 barrier in \textbf{R2} is actually lower than that in \textbf{R3}, 0.7 and 16.0 kJ mol$^{-1}$, respectively (see Figure \ref{fig:RX_CCHonW18}). This is because, in the former, TS1 is more reactant-like than in the latter due to the fact that, in the R structure of \textbf{R2}, the reacting water molecule is not fully saturated by H-bonds (i.e., 2 H-bonds as donor + 1 H-bond as acceptor) while this is the case in \textbf{R3} (i.e., 2 H-bonds as donor + 2 H-bonds as acceptor). The lack of the H-bond in R of the \textbf{R2} reaction allows the water "unsaturated" lone pair to form the hemibonded system with CCH, in which the C-O bond is half-formed. Thus, to reach the TS1 structure, no significant energy requirements and geometrical changes are needed, and hence the low energy barrier. In contrast, in \textbf{R3}, to form TS1 from the R structure, the reacting water molecule has to break one of the H-bonds to form the newly C-O bond, implying a more energetic cost and relevant geometrical changes. Nevertheless, the associated quantum tunnelling crossover temperature is rather high, of 244 K, indicating that this channel could be relevant at the interstellar temperatures.

\subsubsection{On the W33 ASW ice cluster model}

\begin{figure*}[!htbp]
    \centering
    \begin{tabular}{@{} c}
    \includegraphics[width=0.7\textwidth]{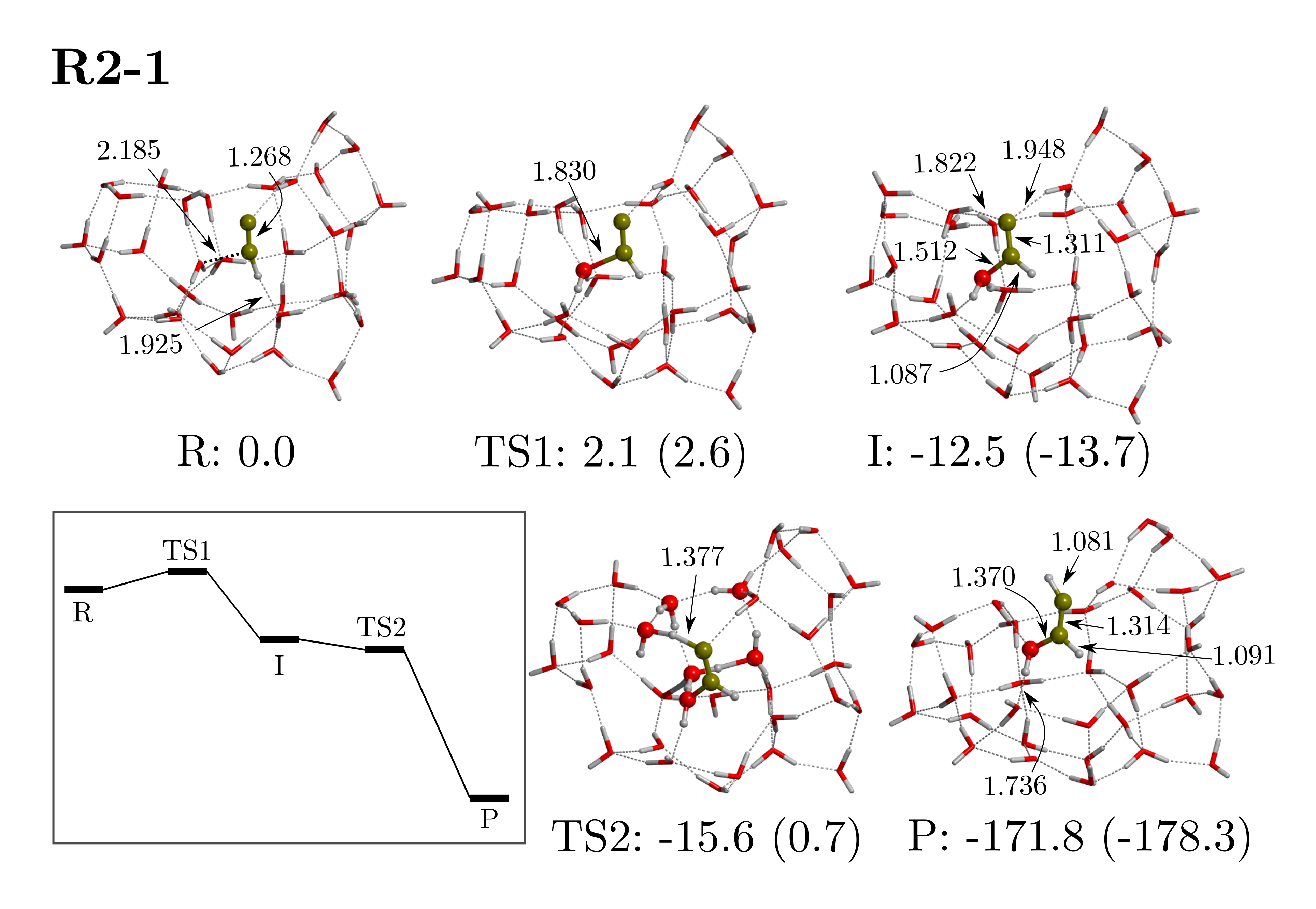} \\
    \hline
    \includegraphics[width=0.7\textwidth]{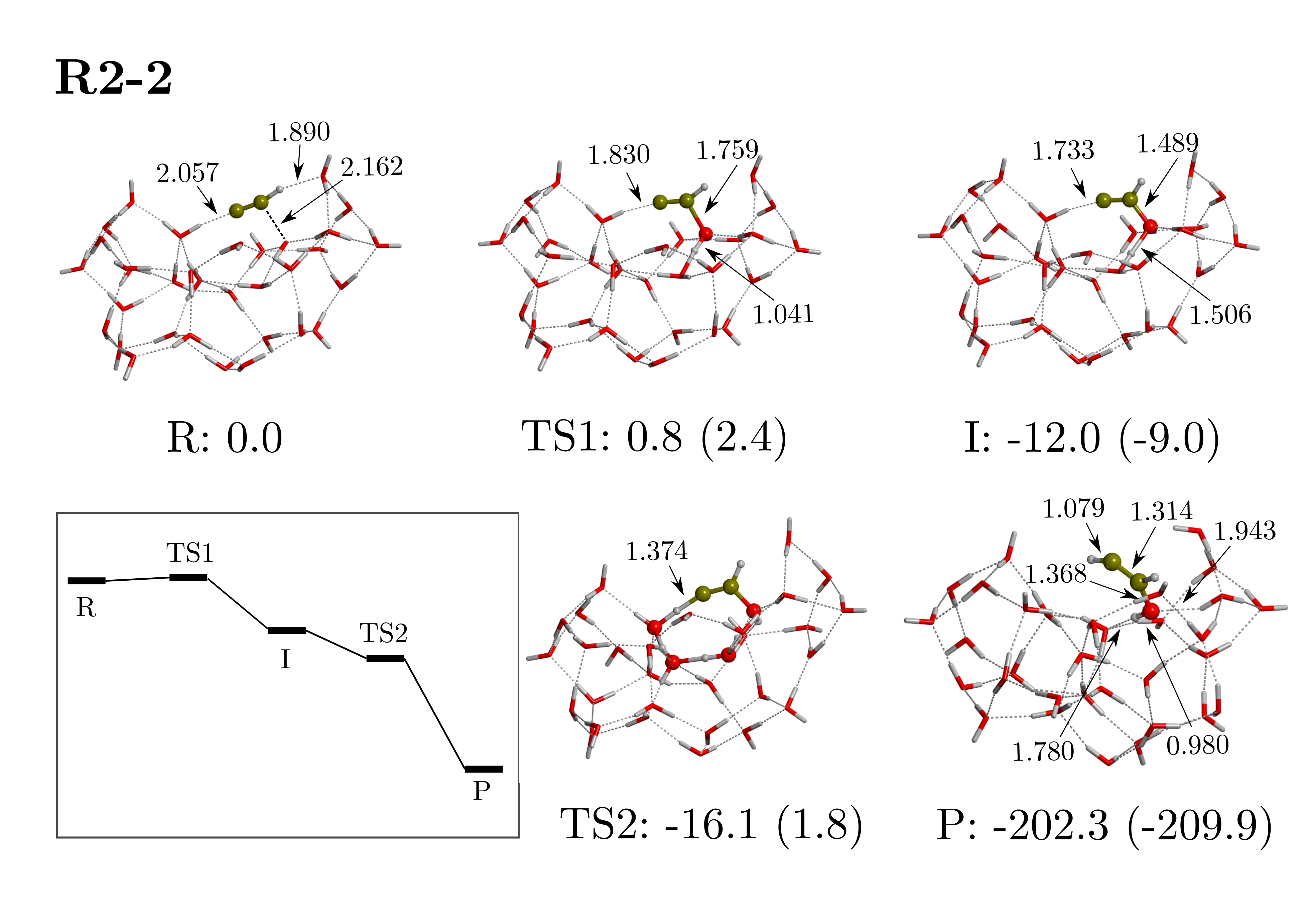}\\
    \hline
    \end{tabular}
    \caption{Computed potential energy surfaces (PESs) of the reactions \textbf{R2-1} and \textbf{R2-2} between CCH and W33 ASW ice cluster model. Bare energy values correspond to relative ZPE-corrected values, while values in parenthesis to those missing this correction. The miniature panels sketch the ZPE-corrected PESs. Energy units are in kJ mol$^{-1}$ and distances in \r{A}.}
    
    \label{fig:RX_CCHonW33}
\end{figure*}

After modeling the reaction on the flat surface of W18, we set out to investigate the possible effects of a cavity structure like that of W33. We tried to reproduce the same three reaction paths as those taking place on W18 but we only found two mechanisms, both similar to \textbf{R2}, hence the name \textbf{R2--1} and \textbf{R2--2}. The stationary points and the energy profiles of these reaction pathways are shown in Figure \ref{fig:RX_CCHonW33}. It is worth mentioning that, to model the reaction path \textbf{R2--2} it was necessary to fix the position of some of the oxygen atoms placed at the edge of the model, since full geometry relaxation lead to the collapse of the cavity. 

In both cases, CCH is located in the cavity, with the central C atom interacting with the oxygen atom of the reactant water molecule of the surface, forming the aforementioned hemibonded systems (see SI for spin densities). Along the reactions, these structures evolve to form the CC(H)-OH$_2$ intermediate. For both cases, this step has a very low ZPE-corrected energy barrier (2.1 and 0.8 kJ mol$^{-1}$ for TS1 and TS2, respectively). The second step involves, for both paths, the proton transfer from the OH$_2$ moiety to the terminal C atom forming the HCCHOH radical, also by means of a water-assisted proton-transfer mechanism. The two paths lead to the same radical because the H-bond network enables that the water-assisted proton-transfer connects with the terminal C atom and not the central one. These second steps are energetically below the corresponding intermediates (by including ZPE-corrections) and accordingly the water-assisted proton-transfers proceed in a barrierless fashion. Since the reaction energies are very large and negative, the computed reaction on W33 are energetically very favourable, similarly to the \textbf{R2} reaction occurring on W18. 


\subsection{Towards the formation of ethanol: hydrogenation of vinyl alcohol }

As shown above, reaction of CCH with an icy water molecule leads to the formation of CHCHOH and H$_2$CCOH.
From these species, to reach ethanol, a set of hydrogenation steps are necessary, as sketched in reactions \ref{eqn:vinyl_a_1}--\ref{eqn:vinyl_a_3}. 

\begin{equation}
\text{HCCHOH/H$_2$CCOH + H $\to$ CH$_2$CHOH}
\label{eqn:vinyl_a_1}
\end{equation}

\begin{equation}
\text{CH$_2$CHOH + H $\to$ CH$_3$CHOH/CH$_2$CH$_2$OH}
\label{eqn:vinyl_a_2}
\end{equation}

\begin{equation}
\text{CH$_3$CHOH/CH$_2$CH$_2$OH + H $\to$ CH$_3$CH$_2$OH}
\label{eqn:vinyl_a_3}
\end{equation}

The first hydrogenation forms vinyl alcohol (VA). As this H-addition is a radical-radical coupling, we consider it as barrierless. The second hydrogenation step is the H-addition to the VA. It involves the reactivity between a closed-shell species (namely, VA) with a radical (H) and, accordingly, it is expected to have an activation barrier. Interestingly, depending on the C in which the H-addition takes place, the species formed is either CH$_3$CHOH or CH$_2$CH$_2$OH. The third and final hydrogenation step leads to the formation of ethanol, irrespective of the initial radical species. This H-addition is, again, a radical-radical coupling and accordingly it is expected to be barrierless in a similar fashion as the first hydrogenation. According to this reactive scheme, to investigate on ethanol formation, we simulated only the reaction \ref{eqn:vinyl_a_2} on the W18 ice model. 

\begin{figure*}[!htb]
    \begin{tabular}{@{} c}
    \includegraphics[width=0.6\textwidth]{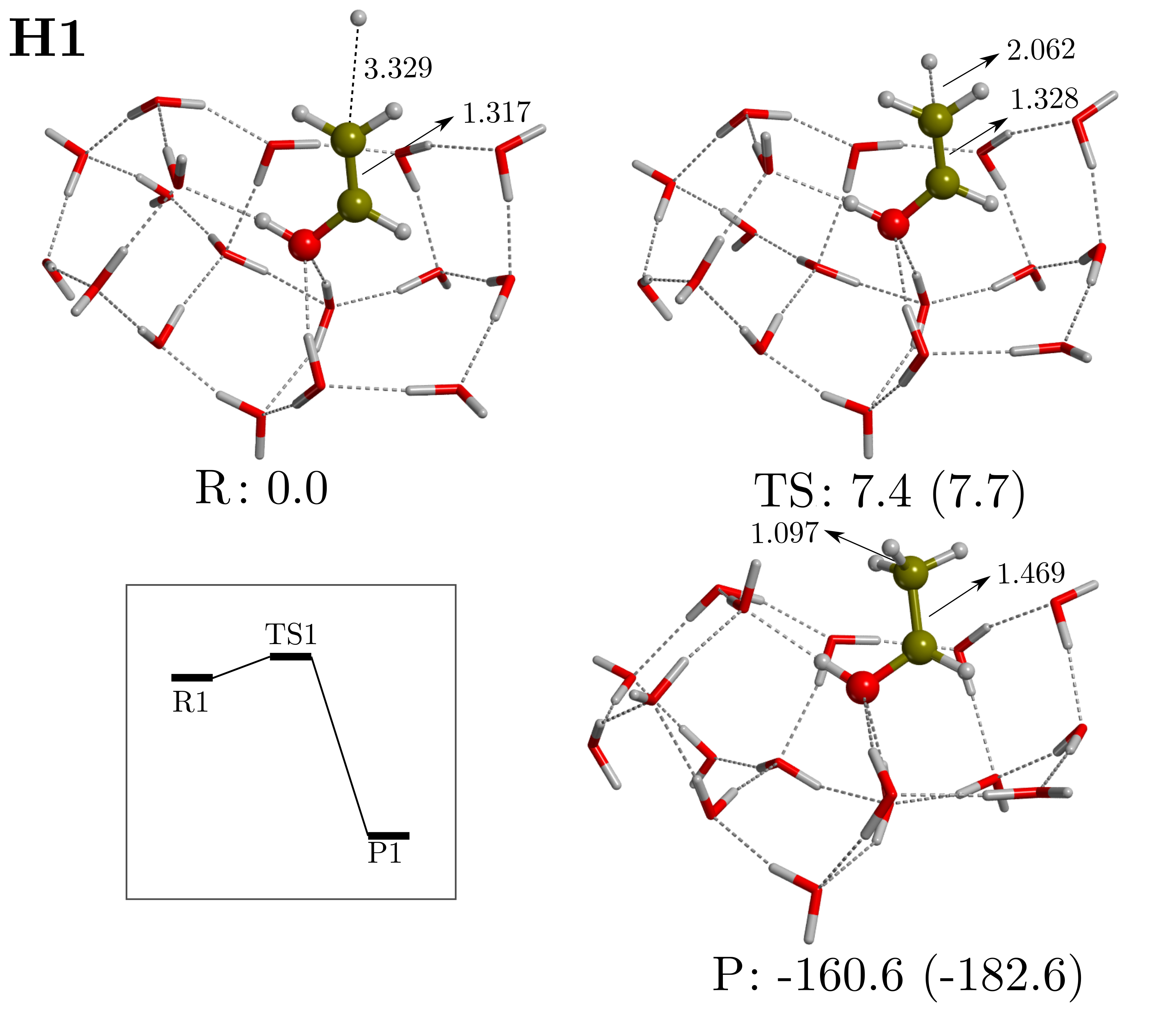} \\
    \hline
    \includegraphics[width=0.6\textwidth]{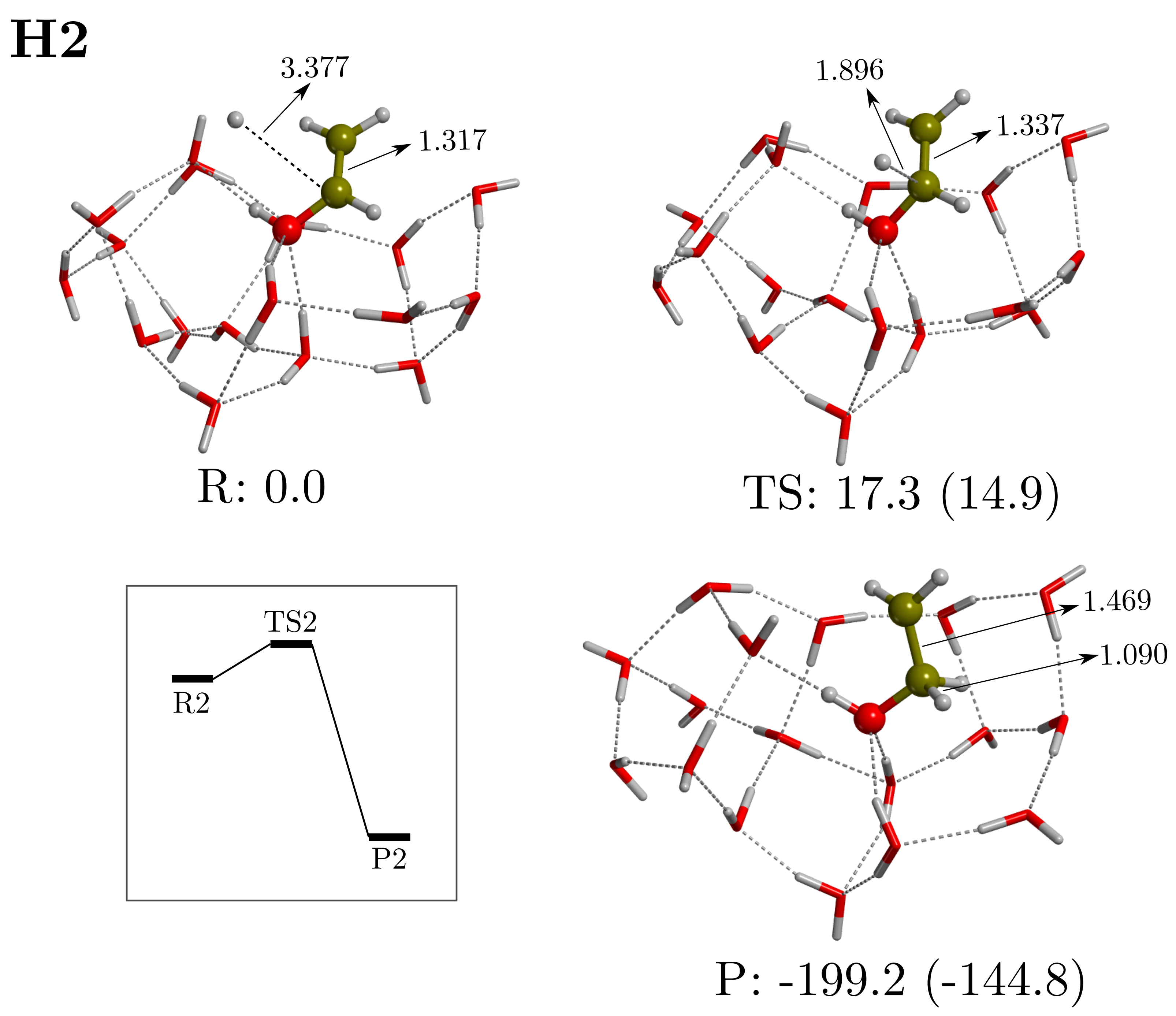}\\
    \hline
    \end{tabular}
       \caption{Computed potential energy surfaces (PESs) of the hydrogenation reactions \textbf{H1} and \textbf{H2} on W18 ASW ice cluster model. Bare energy values correspond to relative ZPE-corrected values, while values in parenthesis to those missing this correction. The miniature panels sketch the ZPE-corrected PESs. Energy units are in kJ mol$^{-1}$ and distances in \r{A}.}
    
    \label{fig:RX_VAHonW18}
\end{figure*}

We identified two reaction paths, named as \textbf{H1} and \textbf{H2} (see Figure \ref{fig:RX_VAHonW18}), the difference of which being the C atom that undergoes the H-addition, in analogy to the gas-phase \textbf{H'1} and \textbf{H'2} reactions (see Figure \ref{fig:benchmark_pes}). Both reactions start from a pre-reactant structure in which the H atom is at ca. 3.3 \r{A} from the VA due to the weak interactions between the two partners.
The computed potential energy surfaces indicate that the two hydrogenation reactions present a non-negligible energy barrier. In agreement with gas-phase results, the path leading to the formation of CH$_3$CHOH is more favourable than that resulting with CH$_2$CH$_2$OH, both in terms of energy barriers and reaction energies. Remarkably, the computed energy barriers on W18 are higher (3--4 kJ mol$^{-1}$) than those obtained in the gas-phase. This is probably because the adsorption of VA with the surface induces an enhanced stabilization of the former due to the intermolecular forces between VA and the icy surface. Thus, chemically strictly speaking, the surface does not act as a chemical catalyst but slightly inhibits the process. However, it is worth highlighting that the \textbf{H1} reaction presents a very low potential energy barrier and, because of the involvement of an H atom and the very low temperatures of the ISM, tunneling effects can operate as well. Indeed, its associated quantum tunnelling crossover temperature is 118 K. For reaction \textbf{H2}, the crossover temperature is slightly higher, 174 K, therefore quantum tunneling could also play a role. 

\subsection{Isomerization between vinyl alcohol and acetaldehyde}

Vinyl alcohol and acetaldehyde are tautomers (i.e. structural isomers), and therefore we have here considered the conversion from one to the other according to reaction \ref{eqn:tautomerization}. This isomerization reaction has been computed at MPWB1K-D3(BJ)//6-311++G(2df,2pd)//MPWB1K-D3(BJ)//6-311++G(d,p) for consistency with the methodology applied to the hydrogenation of vinyl alcohol.

\begin{equation}
\text{CH$_2$CHOH $\Longleftrightarrow$ CH$_3$CHO}
\label{eqn:tautomerization}
\end{equation}

Among them, acetaldehyde is the most stable isomer (by $\sim$41 kJ mol$^{-1}$). In the gas phase, the barrier connecting vinyl alcohol with acetaldehyde is very high (237.9 kJ mol$^{-1}$, see SI). This is because the mechanism involves an intramolecular H-transfer from the OH of vinyl alcohol to the terminal C atom, with a highly strained transition state.

\begin{figure*}[!htb]
    \begin{tabular}{@{} c}
    \includegraphics[width=0.6\textwidth]{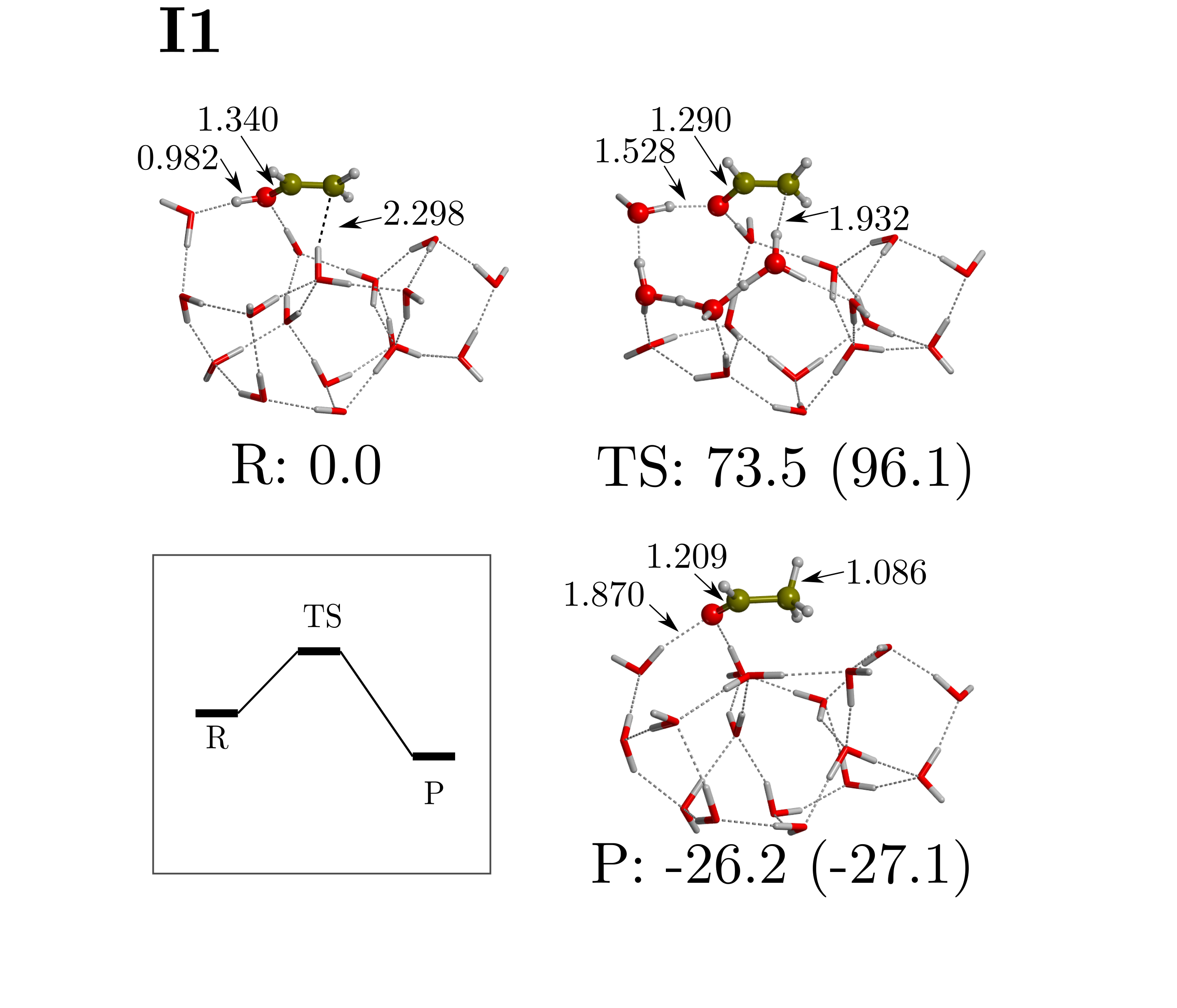} \\
    \hline
     \includegraphics[width=0.6\textwidth]{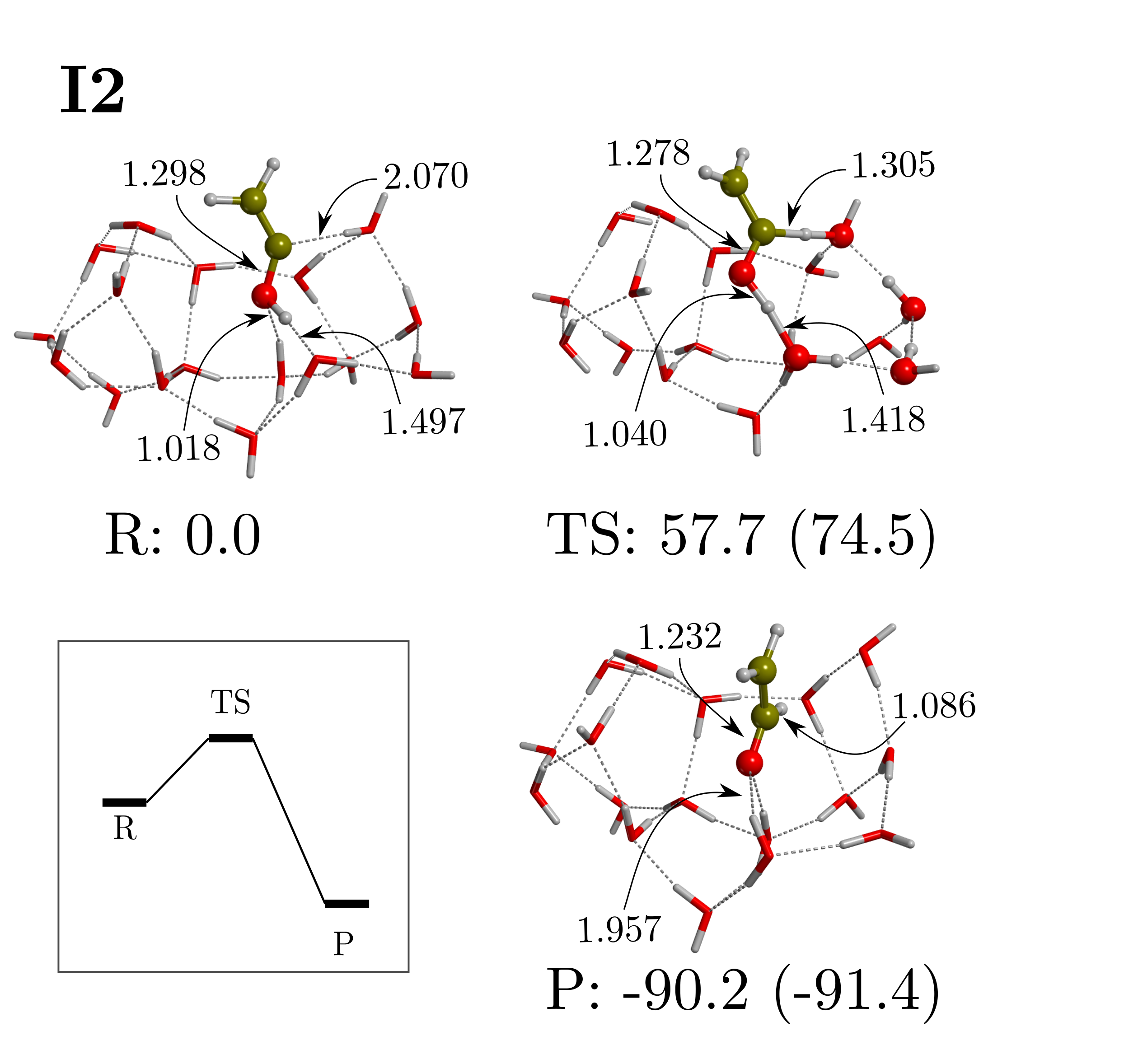} \\
    \hline
    \end{tabular}
    \caption{Computed potential energy surfaces (PESs) of the isomerization between vinyl alcohol and acetaldehyde \textbf{(I1)} and the analogue reaction involving their precursor, i.e., the product P2 of R3 \textbf{(I2)} on W18 ice cluster model at MPWB1K-D3(BJ)/6-311++G(2df,2pd)//MPWB1K-D3(BJ)/6-311++G(d,p). Bare energy values correspond to relative ZPE-corrected values, while values in parenthesis to those missing this correction. The miniature panel sketches the ZPE-corrected PESs. Energy units are in kJ mol$^{-1}$ and distances in \r{A}.}
    \label{fig:va_to_ac}
\end{figure*}

If we now consider this reaction to take place through the water molecules of W18, as shown in Figure \ref{fig:va_to_ac}, we can see that a water-assisted proton-transfer mechanism can take place, lowering the activation energy barrier down to 73.5 kJ mol$^{-1}$ when the reaction starts from vinyl alcohol (I1), and to 57.7 kJ mol$^{-1}$ when it starts from its less hydrogenated precursor (I2). The latter reaction produces CH$_2$CHO, which can be successively hydrogenated to form acetaldehyde. The difference in the two activation barriers is probably due to the structure of the ring through which the proton is transferred, that is, in I1 it is more strained. This is a great example of how interstellar ices act as chemical catalysts. However, computed energy barriers are very high to be surmountable under interstellar conditions and accordingly these channels seem unlikely. Nevertheless, we would like to point out that the barrier of I2 is narrower and lower than that of I1 (\citep[with barriers widths of about 1.6 and 2.2 \r{A}, respectively, assuming the asymmetric Eckart barrier model, see][]{Peters2011_eckart}. Therefore, I2 would be the most efficient mechanism among the two, if assuming that quantum tunneling does play a role.


\section{Discussion and Astrophysical Implications}\label{sec:discuss}

CCH is a highly reactive species that easily reacts with water. As it is a C-centered radical and contains both donor and acceptor H-bond atoms (the H and the C-end atoms, respectively), it tends to form H-bonded and hemibonded complexes with water ice molecules. 
The most relevant computed energetics of CCH reactivity on our ASW ice models are summarized in Table \ref{tab:astronomers}. On W18, \textbf{R3} presents an activation barrier of 14.4 kJ mol$^{-1}$ and, accordingly, it is not \textit{a priori} an efficient path to form vinyl alcohol, although it could still be relevant if  quantum tunnelling effects work. The first step of \textbf{R1} is barrierless to form HCCH + OH but the reaction stops at this stage. This is because the second step has an intrinsic energy barrier of 34.6 kJ mol$^{-1}$ (the energy difference between TS2 and I, see Figure \ref{fig:RX_CCHonW18}). Moreover, the energy released by the first reaction step will no longer be available for the second step, as water ices tend to efficiently dissipate chemical energy fast \cite{pantaleone2020,pantaleone2021}. Therefore, \textbf{R1} will lead to the formation of a highly stabilized acetylene, which will hardly react with OH remaining stuck on the ice. However, this reaction can be an effective channel towards the formation of OH radicals on the ice surfaces without the need of a direct energy processing. This can be of importance because the generated OH could participate in further surface reactions in the form of OH additions. 
\textbf{R2}, at variance with the other two reactions, is an effective barrierless reaction path towards vinyl alcohol formation. Because of that, this reaction has also been modeled on W33 to study the effects of the cavity on the reactivity. On W33, we found that the reactions are no more barrierless, although the two identified paths present low activation barriers (2.1 and 0.8 kJ mol$^{-1}$ for \textbf{R2-1} and \textbf{R2-2}, respectively). Because these two pathways are equivalent to \textbf{R2} on W18, the effect of the cavity is almost insignificant for the energetics of the path. Therefore, we can consider that the mechanistic steps represented by the \textbf{R2}, \textbf{R2--1} and \textbf{R2--2} pathways constitute the most likely channel to form vinyl alcohol.

The hydrogenation of vinyl alcohol on grains leads to the formation of ethanol, as occurring in methanol formation from CO \cite{hidaka2009,watanabe2002,osamura2004,nagaoka2005,fuchs2009} and ethane formation from C$_2$H$_2$ and C$_2$H$_4$ \cite{Bennett-Mile_1973,Hiraoka_2000, Kobayashi_2017}. The H-addition to vinyl alcohol is the only hydrogenation step not involving a barrierless radical-radical coupling and presents an energy barrier of 7.4 and 17.3 kJ mol$^{-1}$ for the \textbf{H1} and \textbf{H2} channel (see Table \ref{tab:astronomers}). Thus, \textbf{H1} is the most energetically favoured pathway for the formation of the ethanol radical precursor. Despite the computed relatively high energy barriers considering very low temperatures, these reactions can proceed through tunneling and accordingly operative in the reaction chain starting from the adsorption of CCH on water ice.

These quantum chemical results can partly be related to the experimental findings mentioned in the \S~\nameref{sec:intro}. Indeed, part of the experimental synthetic routes have been simulated here, providing an atomistic interpretation (including the energetics) for the formation of vinyl alcohol followed by it hydrogenation to form ethanol. The main difference between the experiments and our computations is that, in the experiments, the C$_2$H$_2$/H$_2$O ices need to be processed to trigger reactivity (probably due to generating the radical reactive species like CCH and OH), while in our simulations the assumption is that the reaction does not require energetic processing of ices, since the CCH is readily available in the gas phase in a wide variety of environments.

\begin{table}[!htb]
\centering
\caption{Summary of the energetics of the simulated reactions and their products.}
\label{tab:astronomers}
\resizebox{\columnwidth}{!}{%
\begin{tabular}{|l|c|c|c@{}|}
\hline
\multicolumn{2}{|c|}{Reaction }         & Barrier & Product  \\
\hline
\multirow{3}{*}{W18} &  R1  & NO  &  HCCH + OH \\
 & R2 & NO  & HCCHOH   \\
 & R3 & 14.4 kJ mol$^{-1}$  & H$_2$CCOH/HCCHOH   \\
\hline
\hline
\multirow{2}{*}{W33} &  R2-1  & 2.1 kJ mol$^{-1}$ &  HCCHOH \\
 & R2-2 & 0.8 kJ mol$^{-1}$ & HCCHOH   \\
\hline
\hline
\multirow{2}{*}{W18} &  H1  & 7.4 kJ mol$^{-1}$ &  CH$_3$CH$_2$OH \\
 & H2 & 17.3 kJ mol$^{-1}$ & CH$_3$CH$_2$OH   \\
 \hline
\end{tabular}%
}
\end{table}

We also considered the isomerization of vinyl alcohol into acetaldehyde (and the same for their less hydrogenated precursors), since in the experiments these reactions were suggested to explain the presence of both vinyl alcohol and acetaldehyde. According to our calculations, however, these reactions, although being catalyzed by the surfaces thanks to a water-assisted proton-transfer mechanism, present high energy barriers, rendering them poorly competitive to the final H additions. However, experimental authors pointed out that the isomerization could take place thanks to the exothermicity of the previous steps, in which the energy released along the reaction steps can be used to overcome the isomerization energy barrier. Our computed energetic data is consistent with this view. The very favorable reaction energies shown by the reactions make that the energy barriers of the isomerization processes lay below the pre-reactive asymptotic states and therefore, they can be overcome by making use of the nascent reaction energies. However, one should bear in mind that water ice surfaces are extraordinary third bodies \cite{pantaleone2020,pantaleone2021} and accordingly, the direct transfer of the previous reaction energies to surmount the isomerization energy barriers is doubtful. To shed some light onto this aspect, dedicated \textit{ab initio} molecular dynamics simulations are compulsory, which is out of the scope of the present work. 

Finally, results presented in this work are very relevant in the framework of cold Astrochemistry. 
The presence of abundant ($\sim 10^{-9}--10^{-8}$) gaseous CCH radicals in cold ($\leq 20$ K) regions, where water ices envelope the refractory cores of the interstellar dust grains (see \S~\nameref{sec:intro}, can lead to the formation of vinyl alcohol and ethanol, in addition to HCCH and OH through a competitive reaction channel. 
Remarkably, at variance with most of the experimentally proposed mechanisms (see \S~\nameref{sec:intro}), the formation of the aforementioned products does not require the energetic processing of interstellar ices. 
Regarding the formation of iCOMs, our proposal complements the non-energetic reaction scheme of \citet{Chuang2020_C2H2_O2}, in which OH radicals attack C$_2$H$_2$ to form iCOMs. 
We want to point out that CCH (and the other intermediate radicals) could also be destroyed by other competitive surface reactions, e.g., by H-abstraction reactions with H$_2$, or H-additions. This will be taken care of in the future.

More in general, the astrochemical processes in cold regions such as the ones described in this work are important to improve our understanding of the presence of complex species detected in the cold ($<$20 K) outskirts of prestellar cores during the last decade (e.g. \cite{Bacman2012, Cernicharo2012, Vastel2014, Jaber2014, Jimenez-Serra2016, Scibelli2020}). 
In this vein, ethanol has recently gained some attention, as it has been advocated to be the precursor of several iCOMs formed by cold gas-phase reactions (the genealogical tree of ethanol \cite{skouteris2018, vazart2019}). 
Indeed, the correlation of glycoladehyde and acetaldehyde abundances observed towards a number of interstellar sources has been shown to follow very well the theoretical predictions when their synthesis takes place through this gas-phase scheme \cite{vazart2020}.
In the present work, we showed that an efficient paths exists for the formation of ethanol on the surfaces of interstellar ices. 
However, the non-thermal desorption of ethanol (and of the other products) remains as a crucial missing step in the sequential events linking the chemistry of iCOMs occurring on the surface of grains and in the gas phase in cold regions, this issue undoubtedly being a central matter of further investigation.

\section{Conclusions} \label{sec:conclusions}

Interstellar complex organic molecules (iCOMs) have been detected in different astrophysical environments. However, the chemistry leading to their formation is not unambiguously known. Two prevailing paradigms have been largely used to rationalize their presence in the interstellar medium: one advocating reaction in the gas-phase, the other on the surfaces of icy grains. In this work, we focused on the latter by computing the reaction of CCH with a H$_2$O molecule forming part of the ice structure, leading to vinyl alcohol (CH$_2$CHOH), which upon hydrogenation is converted into ethanol (CH$_3$CH$_2$OH). This reaction is proposed as an alternative synthetic route for iCOMs beyond the commonly assumed radical-radical couplings. Investigations have been performed by means of DFT quantum chemical simulations and adopting cluster models of 18 and 33 water molecules (W18 and W33) to mimic the icy surfaces.  
For the reaction of CCH with H$_2$O, three different reaction pathways have been elucidated, leading to the formation of HCCH + OH, CHCHOH and H$_2$CCOH. Some cases present small activation barriers but in others the reactions are barrierless when zero-point energy corrections are accounted for. Hydrogenation of vinyl alcohol on the W18 cluster has been found to present activation barriers of a certain significance but, for these reactions, quantum tunneling is likely to be at work, speeding them up. Isomerization between vinyl alcohol and acetaldehyde have also been simulated on W18, results indicating that, despite the strong catalytic role played by the water ice, they have a barrier of significant height. Additionally, the direct H-abstraction from the water molecule to CCH, leading to the formation of HCCH and the OH radical on the surface has been found to be an energetically competitive channel. It is worth mentioning that a chemical kinetics treatment of these results are underway in order to compute the rate constants (including tunneling effects explicitly) for each of the proposed reactive channels and hence elucidate branching ratios and formation efficiencies of the simulated paths. 

In summary, results from our calculations indicate that the reaction of CCH with water ice can lead to the formation of vinyl alcohol and, lately, to the production of ethanol, and likely acetaldehyde, without the need of ice energy processing. This conclusion is of relevance in the context of iCOM formation because, according to the genealogical tree of ethanol, this species is the parent molecule through which different iCOMs (e.g., formic acid, formaldehyde, glycolaldehyde, and others ) can form by means of gas-phase processes \cite{skouteris2018,vazart2020}. Thus, in this work, we provide a quantum chemical evidence on the feasibility of our mechanistic proposal, in which ethanol can be formed on interstellar icy grain surfaces, hence linking the two paradigms in the synthesis of iCOMs, at least in this particular case. The missing link between on-grain and gas-phase chemistry stands in the non-thermal desorption of ethanol and its precursors, which should be a subject for further investigation.

\begin{acknowledgement}

This project has received funding within the European Union’s Horizon 2020 research and innovation programme from the European Research Council (ERC) for the projects ``The Dawn of Organic Chemistry” (DOC), grant agreement No 741002 and ``Quantum Chemistry on Interstellar Grains” (QUANTUMGRAIN), grant agreement No 865657.
The authors acknowledge funding from
the European Union’s Horizon 2020 research and innovation program Marie Sklodowska-Curie for the project ``Astro-Chemical Origins” (ACO), grant agreement No 811312.  
AR is indebted to ``Ram{\'o}n y Cajal" program. 
MINECO (project CTQ2017-89132-P) and DIUE (project 2017SGR1323) are acknowledged. 
Finally, we thank Prof. Gretobape for fruitful and stimulating discussions.

Most of the quantum chemical calculations presented in this paper were performed using the GRICAD infrastructure (https://gricad.univ-grenoble-alpes.fr), which is partly supported by the Equip@Meso project (reference ANR-10-EQPX-29-01) of the programme Investissements d'Avenir supervised by the Agence Nationale pour la Recherche. Additionally, this work was granted access to the HPC resources of IDRIS under the allocation 2019-A0060810797 attributed by GENCI (Grand Equipement National de Calcul Intensif). CSUC supercomputing center is acknowledged for allowance of computer resources.\\
\end{acknowledgement}



\begin{suppinfo}

The following files are available free of charge:
\begin{itemize}
  \item Structures and errors of the benchmarking study; energetics relative to the CCH + \ch{H2O} reactions on W18 and W33; contributions to the binding energies of CCH on W18 and W33; Mulliken spin densities and spin density maps of the complexes CCH + \ch{H2O} in the gas phase and on W18 and W33; energetics relative to the hydrogenation of vinyl alcohol; energetics relative to the isomerization of vinyl alcohol into acetaldehyde,
\end{itemize}

\end{suppinfo}

\author{
  Jessica Perrero \orcid{0000-0003-2161-9120} \and
  Joan Enrique-Romero \orcid{0000-0002-2147-7735} \and
  Berta Mart\'{i}nez-Bachs \orcid{0000-0002-1290-7019} \and
  Cecilia Ceccarelli \orcid{0000-0001-9664-6292} \and
  Nadia Balucani \orcid{0000-0001-5121-5683} \and
  Piero Ugliengo \orcid{0000-0001-8886-9832} \and
  Albert Rimola \orcid{0000-0002-9637-4554} 
}
\onecolumn
\bibliography{mybib}

\end{document}